\documentclass{aastex701}
\usepackage{booktabs}

\usepackage{enumitem}
\usepackage{graphicx}
\usepackage{hyperref}
\usepackage{cleveref}

\begin{document}

\title{A Machine-Learning Approach for Identifying CME-Associated Stellar Flares in TESS Observations}

\author[0009-0005-0256-9035,sname='Shi']{Yu Shi}
\affiliation{State Key Laboratory of Public Big Data, Guizhou University, Guiyang 550025, People's Republic of China}
\email{gs.yushi24@gzu.edu.cn}  

\author[0000-0001-8037-5256,sname='Lu']{Hong-Peng Lu} 
\affiliation{State Key Laboratory of Public Big Data, Guizhou University, Guiyang 550025, People's Republic of China}
\affiliation{College of Physics, Guizhou University, Guiyang 550025, People's Republic of China}
\email{hplu@gzu.edu.cn}

\correspondingauthor{Hong-Peng Lu}
\email{hplu@gzu.edu.cn}

\author[0000-0002-2394-9521,sname='Zhang']{Li-Yun Zhang}
\affiliation{College of Physics, Guizhou University, Guiyang 550025, People's Republic of China}
\email{liy_zhang@hotmail.com}

\author[0000-0002-1900-9289,sname='Su']{Tian-Hao Su}
\affiliation{College of Physics, Guizhou University, Guiyang 550025, People's Republic of China}
\email{4828783@qq.com}

\author[0009-0007-7785-5217,sname='Tan']{Chao Tan}
\affiliation{State Key Laboratory of Public Big Data, Guizhou University, Guiyang 550025, People's Republic of China}
\email{gs.ctan24@gzu.edu.cn}

\begin{abstract}
Coronal mass ejections (CMEs) are major drivers of stellar space weather and can strongly influence the habitability of exoplanets. However, compared to the frequent occurrence of white-light flares, confirmed stellar CMEs remain extremely rare. Whether such flares are commonly accompanied by CMEs is a key question for solar–stellar comparative studies. Using Sun-as-a-star soft X-ray flare light curves observed by the GOES XRS 1--8~\AA\ channel, we compiled a sample of 1,766 M-class and larger solar flares and extracted features with both deep convolutional neural networks and manual methods. Five machine-learning classifiers were trained to distinguish eruptive from confined flares, with the random forest model achieving the best performance (true skill statistic; TSS = 0.31). This TSS value indicates that the model possesses a moderate ability to discriminate between eruptive and confined flares. Normalized white-light and GOES XRS flare light curves show broadly consistent temporal evolution, reflecting their shared energy-release history and supporting a probabilistic transfer of the model to white-light flare data. We applied the best-performing RF model to 41,405 TESS-detected flares on FGKM-type main-sequence stars, predicting that approximately 47\% of events show CME-like morphological characteristics, with the model-implied intrinsic association fraction lying in the range 35\%--60\%. Intriguingly, the CME occurrence rate decreases with increasing flare energy, indicating that the most energetic flares may be more strongly confined by overlying magnetic fields. These results provide new insight into flare–CME connections in diverse stellar environments and have important implications for assessing the impact of stellar eruptive activity on exoplanetary atmospheres.
\end{abstract}

\keywords{
\uat{Solar flares}{1496} --- 
\uat{Stellar flares}{1603} --- 
\uat{Solar coronal mass ejections}{310} --- 
\uat{Stellar coronal mass ejections}{1881} --- 
\uat{Machine learning}{1845}
}


\section{Introduction} 
Stellar flares are sudden, intense releases of magnetic energy in stellar atmospheres that emit radiation across the electromagnetic spectrum \citep[e.g.,][]{2024LRSP...21....1K}. Such flares are often accompanied by coronal mass ejections (CMEs)—colossal eruptions of plasma and magnetic fields from the corona that can travel at velocities of hundreds to thousands of~km\,s$^{-1}$ \citep[e.g.,][]{2011LRSP....8....6S,2011LRSP....8....1C,2025MNRAS.536.1089H}. Together, flares and CMEs constitute crucial aspects of stellar activity, affecting not only the star’s own evolution \citep[e.g.,][]{2024ApJ...971..153X} but also the environments of surrounding planets \citep[e.g.,][]{2023FrASS..1064076G}. Energetic flares and CMEs can heat and erode planetary atmospheres, and in extreme cases, frequent large flares and CMEs may lead to complete atmospheric loss for close-in exoplanets, thus significantly impacting their habitability \citep[e.g.,][]{2019LNP...955.....L,2023MNRAS.518.2472R}.

Thanks to space telescopes such as the Kepler mission and the Transiting Exoplanet Survey Satellite (TESS), astronomers have detected an abundance of stellar white-light flares on late-type main-sequence stars of spectral types F, G, K, and M \citep[e.g.,][]{2012Natur.485..478M,2016ApJ...829...23D, 2019ApJS..243...28L,2020AJ....159...60G,2024Sci...386.1301V}. In stark contrast to the prolific flare detections, only a handful of stellar CME candidates have been reported to date \citep[e.g.,][]{2019ApJ...877..105M,2022SerAJ.205....1L,2023ScSnT..53.2021T,2024Univ...10..313V}. These rare CME candidates have been identified primarily through spectral Doppler-shift signatures (transient blue-shifted line asymmetries indicating outward-moving plasma) \citep[e.g.,][]{2019A&A...623A..49V,2022NatAs...6..241N,2023ApJ...953...68L,2025ApJ...978L..32L}, and via coronal dimming observations (drops in X-ray/EUV flux following flares, analogous to solar coronal dimming) \citep[e.g.,][]{2021NatAs...5..697V,2024ApJ...970...60X, 2025ApJ...987L..22C}. Even using these methods, convincing detections of stellar CMEs remain exceedingly scarce. This clear discrepancy between the frequent stellar flares and the rarity of CMEs raises a pivotal question: Do the frequent white-light flares on late-type stars typically have accompanying CMEs, and if so, what fraction of flares are accompanied by CMEs? Addressing this question is key to comparative studies of solar and stellar eruptive activity.

To date, no systematic investigation has predicted stellar CME occurrence based on the properties of stellar flares themselves. Specifically, no prior work has attempted to use information encoded in stellar white-light flare light curves to infer whether a given flare was accompanied by a CME. A recent effort by \citet{2025A&A...694A.161S} endeavored to distinguish eruptive flares (with CMEs) from confined flares (without CMEs) using Sun-as-a-star flare observations in the He~\textsc{ii} 304\,\AA\ channel. By applying principal component analysis to solar flare light curve morphology, they searched for differences between eruptive and confined events—but found no obvious distinguishing features.

On the Sun, the relationship between flares and CMEs has been extensively studied. Statistical surveys show that the CME association rate increases with flare magnitude: around 20\% of C-class flares produce CMEs, while nearly all flares above X3 class are eruptive, with only rare confined exceptions \citep[e.g.,][]{2006ApJ...650L.143Y,2021ApJ...917L..29L}. Whether such flare--CME scaling laws hold for other stars remains an open question. Evidence suggests that active stars may deviate from the solar pattern---for instance, the highly active M dwarf AD Leo exhibits abundant flares but relatively few CMEs, implying solar flare--CME occurrence statistics may not apply directly to such stars \citep{2020A&A...637A..13M}. In parallel, the advent of big data and machine learning has transformed solar flare and CME research in recent years. Machine-learning techniques are now widely employed in solar physics for tasks ranging from flare detection to flare forecasting \citep[e.g.,][]{2023A&A...674A.159D}. However, to date no study has exploited the wealth of solar flare--CME observations to train predictive models and then applied them to Sun-like stars to predict stellar CME occurrence. A key limitation is that stellar flare data are dominated by white-light photometry, whereas solar flares rarely show strong white-light emission \citep[e.g.,][]{2010ApJ...711..185C,2017ApJ...851...91N,2018ApJ...867..159S,2024SoPh..299...11J,2024ApJ...975...69C} leaving too few solar events for effective machine-learning training. To overcome this, Sun-as-a-star soft X-ray observations from the Geostationary Operational Environmental Satellites (GOES) provide an alternative. The GOES X-ray Sensor (XRS) measures full-disk solar flux in the 1--8\,\AA\ band \citep{2024JGRA..12932925W} and has long been used to study solar flare light-curve morphology, offering a rich dataset for training flare-classification models.

In this study, we use GOES XRS 1--8\,\AA\ solar flare light curves to train a machine-learning classifier designed to distinguish eruptive from confined flares, and apply this model to stellar flare data. The classifier, trained on the morphology of Sun-as-a-star solar flare light curves to distinguish eruptive from confined events, is applied to 41,405 white-light flares detected on late-type main-sequence stars by TESS. Our predictions indicate that approximately 47\% of these stellar flares were accompanied by CMEs. Notably, the inferred CME occurrence rate decreases with increasing flare energy---that is, the most energetic stellar flares appear less likely to be associated with CMEs. This trend contrasts with the well-established solar case, yet is consistent with reports for certain active stars where many high-energy flares do not produce CMEs. These findings provide new insight into flare--CME relationships in diverse stellar environments and carry important implications for evaluating the impact of stellar eruptive activity on exoplanetary atmospheres. This paper is organized as follows. Section~\ref{sec:solar} describes the collection and processing of solar flare light curve data. Section~\ref{sec:model} presents the training and evaluation of machine-learning models for distinguishing eruptive from confined flares. Section~\ref{sec:stella} details the search for stellar flares in TESS light curves and the application of the trained model to predict their CME associations. Finally, Section~\ref{sec:summy} provides a brief summary of the work.

\section{Solar Flare Light Curve Data Collection and Processing} \label{sec:solar}
\subsection{Data Sources}
Given the scarcity of solar white-light flare observations and the difficulty in assembling statistically meaningful samples, we constructed the training dataset using soft X-ray light curves obtained from the GOES. Specifically, we selected approximately 4,000 M- and X-class solar flares automatically identified in the 1--8\,\AA\ flux measured by the XRS onboard GOES-8, -10, -15, -16, and -18, covering the period from 1997 to January 2025. The corresponding Sun-as-a-star light curves, with a mean temporal cadence of 1 minute, were extracted for morphological analysis and feature extraction to support the development of the flare–CME classification model.

The GOES data products provide quality flags and documentation\footnote{%
\url{https://data.ngdc.noaa.gov/platforms/solar-space-observing-satellites/goes/goes8/l2/docs/GOES_XRS_Science_Quality_ReadMe.pdf};
\url{https://data.ngdc.noaa.gov/platforms/solar-space-observing-satellites/goes/goes16/l2/docs/GOES-R_XRS_L2_Data_Users_Guide.pdf}
} to support data screening and analysis. To ensure data quality and physical consistency, we excluded events with data gaps, anomalies, or multi-peak structures based on official flags and manual checks. The final sample includes 1,766 high-quality flares (110 X-class and 1,656 M-class), as listed in Table~\ref{tab:solar}. For each event, five additional points before onset and after end were included to assist in background fitting, normalization, and subtraction, improving the robustness of feature extraction.

\begin{deluxetable*}{cccccc}
\tablecaption{Solar Flare Event Catalog\label{tab:solar}}
\tablehead{
\colhead{No.} & \colhead{Start Time (UT)} & \colhead{Peak Time (UT)} & \colhead{End Time (UT)} & \colhead{GOES Class} & \colhead{CME Association}
}
\startdata
1     & 2025/01/31 13:40 & 2025/01/31 14:06 & 2025/01/31 15:10 & M6.7 & 1 \\
2     & 2025/01/29 03:36 & 2025/01/29 04:08 & 2025/01/29 04:38 & M1.0 & 1 \\
3     & 2025/01/28 19:41 & 2025/01/28 19:45 & 2025/01/28 20:44 & M1.7 & 1 \\
4     & 2025/01/27 07:52 & 2025/01/27 08:12 & 2025/01/27 09:25 & M2.6 & 0 \\
5     & 2025/01/24 20:48 & 2025/01/24 21:04 & 2025/01/24 21:30 & M2.7 & 0 \\
\ldots & \ldots & \ldots & \ldots & \ldots & \ldots \\
1762  & 1997/09/24 10:41 & 1997/09/24 11:06 & 1997/09/24 11:42 & M3.0 & 1 \\
1763  & 1997/09/24 02:31 & 1997/09/24 02:48 & 1997/09/24 03:15 & M5.9 & 1 \\
1764  & 1997/09/17 11:29 & 1997/09/17 11:43 & 1997/09/17 12:14 & M1.7 & 1 \\
1765  & 1997/09/02 12:19 & 1997/09/02 12:30 & 1997/09/02 13:00 & M1.0 & 0 \\
1766  & 1997/08/29 22:56 & 1997/08/29 23:32 & 1997/08/29 23:45 & M1.4 & 1 \\
\enddata
\tablecomments{
This table presents a catalog of observed solar flare events. Columns are: (1) event index; (2)–(4) flare start, peak, and end times in UT; (5) GOES classification; and (6) CME association flag, where 1 indicates CME association and 0 indicates no association. Only a subset of events is shown here for illustration. The complete table is provided in machine-readable form in the online journal.}
\end{deluxetable*}

In this study, GOES flare events were matched with CME events from the SOHO/LASCO catalog \citep{1995SoPh..162....1D,1995SoPh..162..357B} to assign binary labels (CME-associated or CME-unassociated) for each flare. The association criterion was defined as follows: if a CME occurred between the flare start time and two hours after its peak time, the flare was considered CME-associated and labeled as eruptive; otherwise, it was labeled as confined. Using this definition, we identified 1,043 eruptive flares, accounting for 59\% of the total sample. Among them, the CME association rate is 89.1\% for X-class flares and 57.1\% for M-class flares.

A two-hour time window has been widely adopted in previous solar flare--CME association studies \citep[e.g.,][]{1991AdSpR..11a..25H,1995A&A...304..585H,2017RAA....17....7S}. This interval reflects the typical delay between a flare and the first appearance of its associated CME in the LASCO C2 field of view, which generally ranges from tens of minutes to about two hours depending on the CME speed. A narrower one-hour window may miss slow CMEs \citep[e.g.,][]{2005JGRA..11012S05Y}, whereas a wider three-hour window tends to introduce unrelated events \citep[e.g.,][]{2009IAUS..257..233Y}. Using this criterion, we obtained CME association rates consistent with previous statistical results \citep[e.g.,][]{2006ApJ...650L.143Y,2009IAUS..257..233Y}, confirming the robustness of our classification scheme.

\subsection{Data Preprocessing}
As the raw flare data are unsuitable for direct model analysis, we first performed systematic preprocessing to ensure the reliability and effectiveness of subsequent model training.

Flare sample times were recorded in Coordinated Universal Time (UTC). All light curves were converted to seconds and temporally shifted so that the flare peak corresponds to time zero. To account for variations in flare duration, we applied linear interpolation to determine the times at half-peak intensity and calculated the full width at half maximum (FWHM). Each time series was then rescaled by expressing time in units of FWHM. This peak-alignment and FWHM-normalization method was first introduced by \citet{2014ApJ...797..122D} and later extended for large-scale statistical studies by \citet{2016ApJ...829...23D}. It has since become a standard technique for stellar flare time-series analysis.

To more accurately extract the net signal representative of flare energy release, we performed a linear fit using five data points immediately preceding and following each flare. The resulting fit defines the local background baseline. Subtracting this baseline effectively removes slowly varying background components present during the event, yielding a background-subtracted flare light curve. To reduce amplitude variation across events and improve numerical stability during model training, all net light curves were peak-normalized to the range of 0 to 1, ensuring uniform scaling across samples.

\subsection{Feature Extraction}
\begin{figure*}[ht!]
\plotone{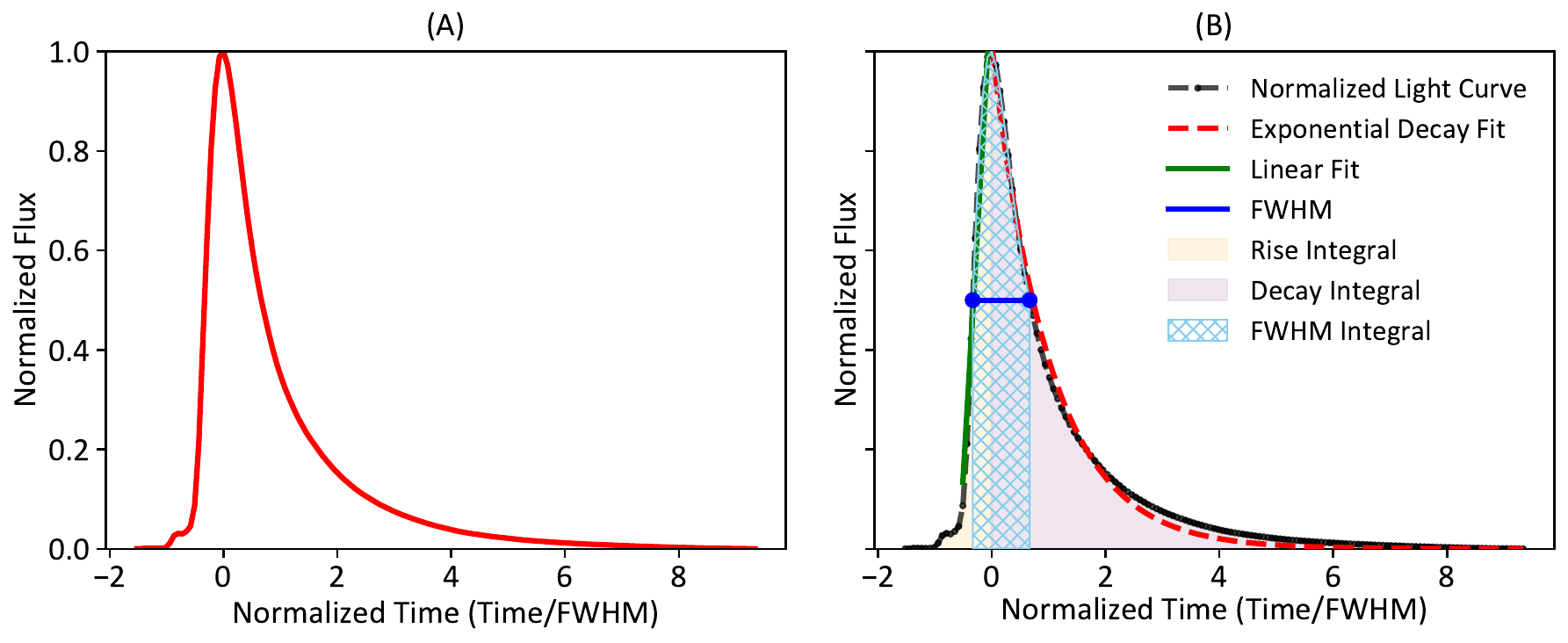}
\caption{Normalization and feature extraction of a solar flare light curve. Panel (A) displays the normalized light curve (red solid line). Panel (B) shows a subset of extracted features, including the normalized curve (black line), a linear fit to the rise phase (green solid line), an exponential decay fit to the decay phase (red dashed line), and the full width at half maximum (FWHM; blue solid line). Shaded areas indicate the integration intervals for the rise phase (light yellow), decay phase (light purple), and FWHM (light blue hatched).}
\label{fig:figure1}
\end{figure*}

As shown in Figure~\ref{fig:figure1}(A), the preprocessed light curves display clearly defined flare profiles. To characterize their temporal evolution, we extracted 33 features: 13 manually defined morphological features and 20 image-based features derived from a deep learning model. Since this study focuses on the morphological characterization of flares, all features were extracted from the preprocessed light curves to accurately represent their structural evolution.

Thirteen manually defined features were extracted to characterize flare dynamics, describing temporal evolution and brightness variations. These include total duration; ratios of rise-to-total and rise-to-decay durations; integrated net fluxes during the rise and decay phases, and their ratio; total integrated flux; integrated flux within the full width at half maximum (FWHM); and ratios of the FWHM integral to the rise, decay, and total integrals, respectively. We also derived the slope from a linear fit to the rise phase and the time constant from an exponential fit to the decay phase. Representative features are illustrated in Figure~\ref{fig:figure1}(B). Collectively, these parameters characterize the structural and temporal properties of flares, facilitating further classification and analysis.

To overcome the limitations of manually defined features, we transformed each normalized flare profile into a fixed-size grayscale image and applied a ResNet50 deep residual network \citep{2016cvpr.confE...1H} to extract high-level morphological features. During preprocessing, axis labels were removed to eliminate redundant visual cues, ensuring the model focuses solely on intrinsic structural and temporal patterns. This approach enhances the generalizability and representational power of the extracted features.

To mitigate overfitting risks from limited sample size and high-dimensional features, the extracted image-based features were reduced to 20 dimensions using principal component analysis (PCA), retaining principal components that explain 80\% of the original variance. These principal components capture distinct aspects of flare morphology, enabling effective dimensionality reduction and noise suppression while enhancing model stability and generalization.

\subsection{Constructing the Training Dataset}
We constructed a classification dataset by concatenating the 13 handcrafted features with 20 image-based PCA features, resulting in 33 fused features. Similar hybrid feature fusion strategies combining handcrafted and CNN-extracted features have been shown effective in various domains \citep[e.g.,][]{SU2020300,KHAN2020105986,VISWANATHAREDDY202023}. To evaluate model robustness under different data partitions, five independent 80\%–20\% training–testing splits with stratified sampling were generated to preserve class distributions, thereby reducing bias and ensuring reliable performance evaluation.

\section{Training and Evaluation of Machine-Learning Models for Eruptive and Confined Flare Classification} \label{sec:model}
\subsection{Model Selection}
This study analyzes 1,766 solar flares of M class and above. As the extracted features are structured numerical vectors and the goal is to distinguish between CME-associated and CME-unassociated flares, the task is formulated as a binary classification problem. Given the moderate feature dimensionality and limited sample size, conventional machine learning methods are well suited for modeling and prediction.

To balance robustness and performance, we evaluated five supervised classifiers: logistic regression (LR), random forest (RF), gradient boosting trees (XGB), linear discriminant analysis (LDA), and linear support vector machine (Linear SVM). Additionally, a dummy classifier was included as a baseline (see Appendix~\ref{sec:A}). These models cover linear and nonlinear, as well as discriminative and ensemble methods, providing a comprehensive evaluation of the discriminative power of the extracted features.

\subsection{Training Methodology}
To effectively prevent data leakage, only the training set distribution was used to normalize the test set during standardization. Specifically, features were scaled using the more outlier-robust RobustScaler. The entire preprocessing and modeling workflow was managed through the Scikit-learn Pipeline, ensuring consistent data transformations across training and testing phases, thereby enhancing process rigor and reproducibility \citep{2011JMLR...12.2825P}.

After initial standardization, classification labels “1” and “0” were assigned to flares associated with CMEs and those without, respectively. To ensure fair comparison among models, a consistent training and evaluation procedure was designed and applied uniformly to all classifiers.

Based on the five train–test splits generated using different random seeds, we independently trained and evaluated each classifier to ensure robust performance estimation. For each split, hyperparameters were optimized using five-fold stratified cross-validation combined with grid search, with classification accuracy as the evaluation metric. The model was then retrained on the full training set using the optimal hyperparameters and evaluated on the corresponding test set. This procedure yielded five independently trained sub-models and their respective test results for each classifier. The final performance metrics were obtained by averaging the test results across all five splits, as summarized in Table~\ref{tab:model_metrics}. This repeated evaluation strategy mitigates biases introduced by any single split and provides a more reliable assessment of model generalization and stability. In addition, class balancing was applied to all classifiers to account for the slight class imbalance in the dataset. For the Random Forest model, we adopted \texttt{class\_weight='balanced\_subsample'}, and analogous weighting schemes were used for the other machine-learning models. Because the imbalance was relatively minor, these standard weighting adjustments were sufficient to mitigate its impact without requiring more aggressive resampling techniques.

Given the binary nature of the classification task, model performance was evaluated using a set of confusion-matrix-based metrics, including accuracy, precision, false discovery rate (FDR), false alarm rate (FAR), and true skill statistic (TSS), as detailed in Appendix~\ref{sec:B}. Since most of these metrics—except TSS—are sensitive to class imbalance \citep[e.g.,][]{2012ApJ...747L..41B,2015ApJ...798..135B}, TSS was selected as the primary evaluation metric to ensure a more balanced and robust assessment of predictive capability.

As shown in Table~\ref{tab:model_metrics}, the RF model attained the highest TSS score of 0.31, substantially outperforming the other five benchmark classifiers. TSS ranges from -1 to 1, with 0 indicating no predictive skill (equivalent to random guessing) and 1 denoting perfect classification. Relative to the baseline model (TSS = 0), the RF model exhibits superior ability to distinguish between CME-associated and unassociated flares, reflecting both reliable discriminatory power and reduced sensitivity to class imbalance. Notably, it also achieves the lowest false positive rate, underscoring its robustness in limiting false alarms. Considering all performance metrics, we select the RF model as the primary classifier for further analysis.

\begin{deluxetable*}{ccccccc}
\tabletypesize{\normalsize}
\tablecaption{Evaluation metrics of different machine learning models \label{tab:model_metrics}}
\tablehead{
\multicolumn{1}{l}{Model} &
\multicolumn{1}{l}{\shortstack{Accuracy\\\footnotesize [0--1]}} &
\multicolumn{1}{l}{\shortstack{Precision\\\footnotesize [0--1]}} &
\multicolumn{1}{l}{\shortstack{F1 Score\\\footnotesize [0--1]}} &
\multicolumn{1}{l}{\shortstack{FDR\\\footnotesize [0--1]}} &
\multicolumn{1}{l}{\shortstack{FPR\\\footnotesize [0--1]}} &
\multicolumn{1}{l}{\shortstack{TSS\\\footnotesize [-1--1]}}
}
\startdata
RF       & $0.64 \pm 0.01$ & $0.75 \pm 0.02$ & $0.66 \pm 0.01$ & $0.25 \pm 0.02$ & $0.28 \pm 0.04$ & $0.31 \pm 0.02$ \\
LR       & $0.63 \pm 0.02$ & $0.71 \pm 0.02$ & $0.67 \pm 0.01$ & $0.29 \pm 0.02$ & $0.37 \pm 0.04$ & $0.26 \pm 0.04$ \\
SVM      & $0.63 \pm 0.01$ & $0.70 \pm 0.01$ & $0.67 \pm 0.01$ & $0.30 \pm 0.01$ & $0.40 \pm 0.02$ & $0.25 \pm 0.02$ \\
XGB      & $0.62 \pm 0.00$ & $0.71 \pm 0.01$ & $0.66 \pm 0.00$ & $0.29 \pm 0.01$ & $0.36 \pm 0.03$ & $0.25 \pm 0.02$ \\
LDA      & $0.63 \pm 0.01$ & $0.64 \pm 0.01$ & $0.73 \pm 0.01$ & $0.36 \pm 0.01$ & $0.67 \pm 0.03$ & $0.17 \pm 0.02$ \\
BaseLine & $0.50 \pm 0.02$ & $0.59 \pm 0.02$ & $0.54 \pm 0.02$ & $0.41 \pm 0.02$ & $0.50 \pm 0.03$ & $0.00 \pm 0.05$ \\
\enddata
\end{deluxetable*}

\subsection{Feature Importance and Physical Interpretation}

Our feature-importance analysis (Appendix~\ref{sec:E}) shows that the deep-learning image features extracted by ResNet50 account for approximately 70\% of the model discriminative power, far exceeding the contribution from manually engineered features. Nevertheless, the most influential manual features provide valuable physical insight into the distinction between eruptive and confined flares. The five highest-ranked manual features are: (1) the time constant from an exponential fit to the decay phase, (2) the ratio of the FWHM integral to the decay-phase integral, (3) the ratio of the FWHM integral to the rise-phase integral, (4) the FWHM integral, and (5) the decay-phase integral.

These features are statistically consistent with well-established differences between solar eruptive and confined flares. Eruptive flares associated with CMEs tend to show longer decay times, stronger gradual-phase energy release, and larger post-peak soft X-ray fluence, reflecting sustained magnetic reconnection in large post-eruption arcades \citep[e.g.,][]{1977ApJ...216..108P, 1992ARA&A..30..113K, 2006ApJ...650L.143Y, 2009IAUS..257..233Y}. Thus, the prominence of the decay time constant and the decay integral indicates that prolonged soft X-ray emission is a strong statistical signature of eruptive behavior, whereas confined flares---whose energy release is restricted by closed magnetic topology---cool more rapidly. Likewise, the two FWHM-to-rise and FWHM-to-decay fluence ratios quantify the relative impulsiveness versus gradual-phase dominance of the flare: eruptive events statistically exhibit larger fractional contributions from the rise and decay phases relative to the peak (FWHM), while confined flares show more sharply peaked, impulsive emission concentrated within the FWHM interval. The high importance of the FWHM integral further reflects the empirical trend that more energetic flares, with larger peak-phase fluence, have a higher probability of producing CMEs.

Taken together, these results demonstrate that the physical features most favored by the model correspond directly to the temporal energy distribution of the flare---its impulsiveness, duration, and gradual-phase radiative output. These trends agree with the established phenomenology of eruptive and confined flares and indicate that our model captures physically meaningful signatures in the GOES soft X-ray profiles when inferring CME association.

\subsection{Extension from the Sun to FGKM-type main-sequence Stars}
To assess the applicability of the model to stellar flare observations in different wavelength regimes, we compared the light curve morphologies of flares in the GOES XRS 1–8~\AA\ broadband channel with those observed in white light. For comparison, we selected the complete white-light flare light curves reported by \citet{2024ApJ...975...69C} that coincide in observation time with our sample and exhibit high signal-to-noise ratios. This cross-band comparison serves to evaluate the model's generalization capability under varying observational modalities and flare emission mechanisms.

\begin{figure*}[ht!]
\plotone{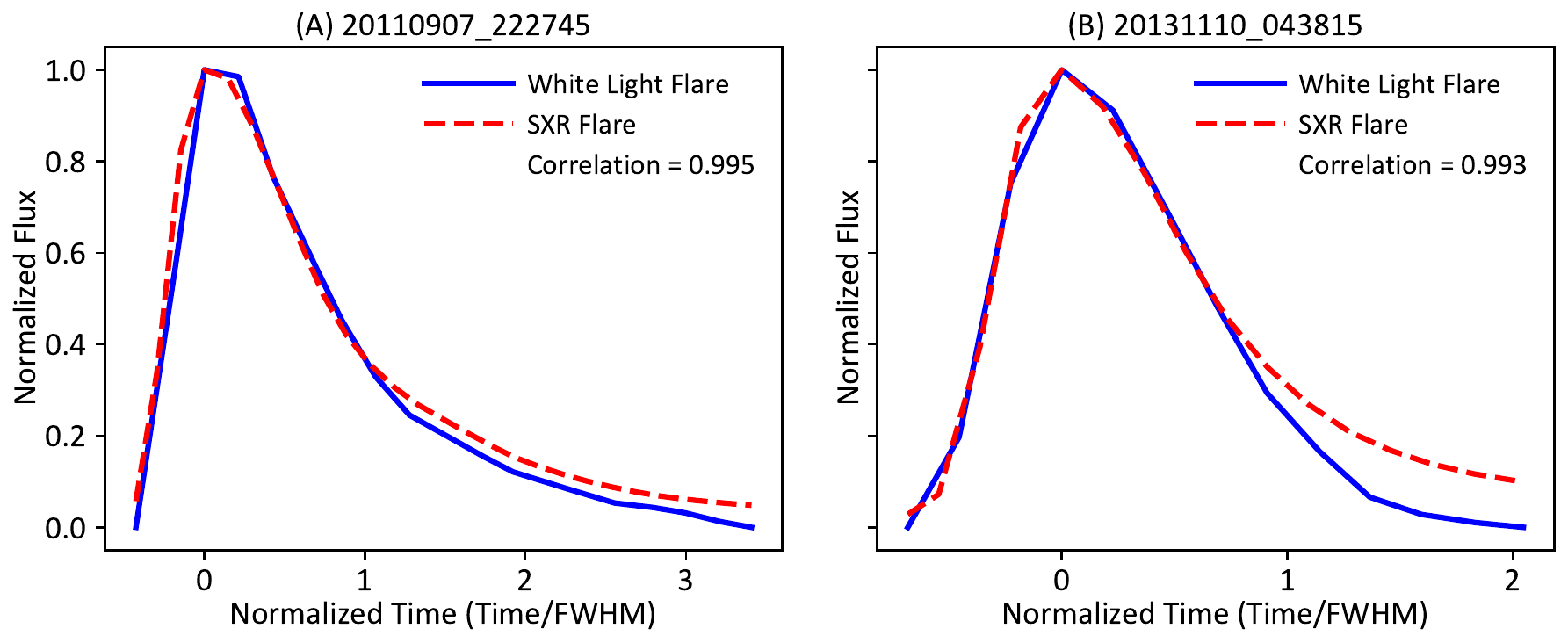}
\caption{Comparison of normalized light curves for white-light flares (solid blue) and the GOES XRS 1–8~\AA\ flares (dashed red). Panels (A) and (B) correspond to events 20110907\_222745 (GOES class X1.8) and 20131110\_043815 (GOES class X1.1), named after the start times of the white-light flares. The x- and y-axes represent normalized time and normalized flux intensity, respectively. Pearson correlation coefficients of 0.995 and 0.993 indicate a strong similarity in light curve profiles.}
\label{fig:fig2}
\end{figure*}

To ensure consistency in cross-band comparison, white-light flares were preprocessed following the same procedure as the GOES XRS 1–8~\AA\ flares. This included time-axis alignment to the flare peak, light-curve normalization, and uniform time-window extraction. Subsequently, linear interpolation was applied to resample the curves, mitigating differences in sampling cadence and enabling direct morphological comparison across wavelengths. As shown in Figure~\ref{fig:fig2}, the standardized profiles of white-light flares and the GOES XRS 1–8~\AA\ flares exhibit strong agreement in overall shape, indicating similar temporal evolution. Beyond visual comparison, we further analyzed the manually extracted features. The results show that solar and stellar flares share similar distributions across multiple feature dimensions and occupy a largely overlapping feature space (see Appendix~\ref{sec:E}). These findings provide empirical and methodological support for transferring models trained on the GOES XRS 1–8~\AA\ flares to the classification and prediction of stellar white-light events.

Although white-light and soft X-ray flares originate from different atmospheric layers—the former predominantly from the lower chromosphere and upper photosphere, and the latter from hot evaporated coronal plasma—they are physically coupled through the flare energy-transport process. In the standard thick-target scenario, magnetic reconnection in the corona accelerates electrons that heat the lower atmosphere and produce strong optical/UV continuum emission (i.e., white-light enhancement), while simultaneously driving chromospheric evaporation that fills post-flare loops with hot plasma radiating in soft X-rays \citep[e.g.,][]{2017LRSP...14....2B, 2015ApJ...809..104A}. Solar observations further reveal strong correlations between white-light properties and GOES soft X-ray fluxes, as well as Neupert-effect--like timing in which the white-light peak coincides with the peak of the GOES soft X-ray derivative and hard X-ray bursts \citep{2002A&A...382.1070V,2016ApJ...816....6K,2024SoPh..299...11J}. These results indicate that, despite forming in different layers, both passbands encode a similar underlying energy-release history. At the same time, solar surveys show that many strong flares remain confined without producing CMEs \citep{2016SoPh..291.1761H}, implying that flare–CME coupling is not one-to-one. Accordingly, we interpret our results as model-predicted CME association probabilities rather than direct measurements of the CME occurrence rate. This probabilistic interpretation aligns the physical limitations of cross-band flare
morphology with the statistical nature of our machine-learning inference and provides a robust foundation for applying the GOES-trained classifier to white-light flares on FGKM-type main-sequence stars.

\section{Searching for Stellar Flares in TESS Light Curves and Predicting Associated CMEs with the Trained Model} \label{sec:stella}

\subsection{Data Sources and Flare Search}
The stellar flare sample was drawn from the two-minute-cadence TESS photometry \citep{2015JATIS...1a4003R} and retrieved from the TESS-SPOC full-frame–image products hosted at MAST (\citealt{2020RNAAS...4..201C}, DOI: 10.17909/t9-wpz1-8s54). We used PDCSAP light curves produced by the SPOC pipeline \citep{2016SPIE.9913E..3EJ, 2020RNAAS...4..201C}, covering Sectors 1 through 81. To remove low-frequency background trends while preserving rapid flare signals, we applied the detrending method proposed by \citet{2022MNRAS.513.4579I}, which combines moving-window smoothing with local polynomial fitting. This technique is effective for flare detection in light curves affected by planetary transits or rapid stellar rotation. Candidate flares were initially identified using the altaipony package \citep{2021JOSS....6.2845I}.

The goal of this study is not to construct a comprehensive TESS flare catalog, but to identify a subset of stellar flares with temporal morphologies closely resembling those of solar flares for subsequent CME association modeling. To achieve this, we applied a two-stage screening procedure combining automated detection with manual inspection. Events exhibiting artifacts, significant background variability, or complex multi-peak structures were excluded to ensure the reliability and consistency of flare profiles. Detailed screening criteria and procedures are described in Appendix~\ref{sec:C}.

Note that the initial stellar flare sample included events from stars off the main sequence, whose flare mechanisms may differ from that of the Sun. Following \citet{2018A&A...616A..10G}, we classified stellar types using color indices (see Appendix~\ref{sec:D}) and retained only flares from FGKM-type main-sequence stars. This filtering yielded 41,405 high-confidence flares from 12,486 stars, with a very low false positive rate. Each flare was processed into the same standardized format as solar flares, including five time steps before and after the flare peak to facilitate background fitting, normalization, and subtraction.

\subsection{Data Preprocessing}
The filtered stellar flare sample was further preprocessed using the same temporal procedure applied to solar flares. Although TESS light curves were detrended during initial flare identification, some residual background variation may remain. To correct for this, we used five data points before and after each flare peak to define a background window for linear baseline fitting. Subtracting the fitted baseline yielded the net flare flux, which was then normalized to its peak value. This procedure ensures consistent scaling and temporal preprocessing across both stellar and solar flare datasets.

\subsection{Feature Extraction and Construction of Prediction Dataset}
Morphological features of stellar flares were extracted using the same procedure as for solar flares, yielding 33 quantitative descriptors. In addition, image-based features were generated from standardized light curve plots, with axis labels removed to prevent implicit encoding of physical scales. To ensure consistency in feature space, principal component analysis (PCA) was applied to reduce the dimensionality to 20. This unified preprocessing pipeline enables direct transfer of solar-trained models to stellar flare classification. The final feature sets were compiled into a dataset suitable for CME association modeling, allowing probability estimates of stellar CMEs.

\subsection{Stellar CME Prediction}

Based on Table~\ref{tab:model_metrics}, the RF classifier achieved the highest overall performance. We therefore constructed an ensemble framework by aggregating five independently trained RF submodels. Two integration strategies were implemented: hard voting, which selects the majority-predicted class, and soft voting, which averages the predicted class probabilities. These ensemble methods enhance predictive stability and improve overall robustness.

Using the best-performing RF mode, we derived CME occurrence probabilities for 41,405 main-sequence stellar flares, resulting in a predicted CME-associated fraction of approximately 47\%. To quantify the statistical uncertainty inherent in this model-derived value, we estimated the corresponding range of intrinsic CME association fractions implied by the classifier’s precision and sensitivity. Following \citet{2020arXiv201016061P} and the prevalence-correction formulation of \citet{ROGAN_WALTER}, the intrinsic CME association fraction can be expressed as
\begin{equation}
    f_{\rm CME} = AP \times \frac{PPV}{Se},
\end{equation}
where $AP$ is the fraction of flares predicted as CME-associated ($AP \simeq 0.47$), and $PPV$ and $Se$ are defined by
\begin{equation}
    PPV=\frac{TP}{TP+FP}, \qquad 
    Se=\frac{TP}{TP+FN},
\end{equation}
with $TP$, $FP$, and $FN$ denoting true positives, false positives, and false negatives. Using the performance metrics derived from the solar training sample ($PPV\approx 0.75$, $Se\approx 0.59$), the implied upper bound on the CME association fraction is
\begin{equation}
    f_{{\rm CME},\,{\rm max}} \approx 0.60.
\end{equation}
A conservative lower bound, assuming perfect sensitivity ($PPV\approx 0.75$, $Se=1$), is
\begin{equation}
    f_{{\rm CME},\,{\rm min}} \approx 0.35.
\end{equation}
Thus, while the RF model predicts CME-like morphological characteristics for $\approx 47\%$ of TESS white-light flares, the corresponding intrinsic CME association fraction lies in the range $0.35$--$0.60$, providing a deeper quantification of the model uncertainty.

Because our results rely on model-predicted CME association probabilities rather than direct observational measurements, it is necessary to quantify the uncertainty arising from model dependence. In this work, our primary analysis is based on the RF classifier. To assess the sensitivity of the inferred CME association probabilities to model choice, we further applied three additional, independent machine-learning classifiers---LR, SVM, and XGB---to the same flare sample. As shown in Table~\ref{tab:model_metrics}, all four models achieve $\text{TSS} \geq 0.25$, indicating non-negligible predictive skill in distinguishing eruptive from confined flares and thus constituting plausible members of the candidate model set.

Following the multimodel inference framework of \citet{burnham2002model}, we quantify the uncertainty associated with model dependence by explicitly accounting for two components: (i) within-model uncertainty, reflecting the statistical and calibration uncertainty intrinsic to each individual classifier, and (ii) between-model variability (a direct measure of model dependence), reflecting the dispersion among predictions from different plausible models. The resulting $AP$ estimates and their corrected intervals are: RF $0.47\ (0.35{-}0.60)$, LR $0.55\ (0.39{-}0.61)$, SVM $0.58\ (0.41{-}0.62)$, and XGB $0.49\ (0.35{-}0.57)$. Given the comparable and non-negligible predictive performance of these models, we treat them as equally plausible and adopt equal weights ($w_i = 1/4$) in the multimodel inference. The model-averaged $AP$ is
\begin{equation}
    \hat{\phi}_{\mathrm{MA}} = \sum_i w_i \hat{\phi}_i = 0.523
\end{equation}
where $\hat{\phi}_i$ denotes the $AP$ estimate from the $i$-th model.

The contribution from model dependence is quantified by the between-model variance,
\begin{equation}
    \mathrm{Var}_{\mathrm{model}} = \sum_i w_i (\hat{\phi}_i - \hat{\phi}_{\mathrm{MA}})^2
\end{equation}
which yields a model-dependence standard deviation of $\sigma_{\mathrm{model}} = 0.044$.

To incorporate within-model uncertainty, we conservatively interpret each corrected interval $[L_i, U_i]$ as an effective two-sided 95\% interval for the $AP$ estimate of the $i$-th model. Under a normal approximation, a two-sided 95\% interval corresponds to $\pm 1.96\sigma$, and the standard error is therefore estimated as
\begin{equation}
    \mathrm{SE}_i \approx \frac{U_i - L_i}{2 \times 1.96}
\end{equation}
where $L_i$ and $U_i$ denote the lower and upper bounds of the corrected $AP$ interval for model $i$, respectively. The within-model variance is then approximated as $\mathrm{Var}(\hat{\phi}_i) \approx \mathrm{SE}_i^2$.

Following \citet{burnham2002model}, the total (unconditional) variance is obtained by combining both components,
\begin{equation}
    \mathrm{Var}_{\mathrm{uncond}} = \sum_i w_i \left[ \mathrm{Var}(\hat{\phi}_i) + (\hat{\phi}_i - \hat{\phi}_{\mathrm{MA}})^2 \right]
\end{equation}
yielding an unconditional standard error of $\mathrm{SE}_{\mathrm{uncond}} = 0.073$.

The relative contribution of model dependence to the total uncertainty is quantified as
\begin{equation}
    \frac{\mathrm{Var}_{\mathrm{model}}}{\mathrm{Var}_{\mathrm{uncond}}} \approx 0.37
\end{equation}
indicating that approximately 37\% of the total variance arises from model dependence. Together with the direct measure of between-model dispersion ($\sigma_{\mathrm{model}} = 0.044$, the quantified model-dependent uncertainty), this demonstrates that model choice constitutes a substantial, though not dominant, source of uncertainty in the inferred CME association probabilities.

We further examined the distribution of eruptive and confined flares across different stellar spectral types. The sample includes 2,830 F-type stars with 8,283 flares (CME association rate of 42\%), 1,375 G-type stars with 2,661 flares (45\%), 1,579 K-type stars with 4,302 flares (47\%), and 6,702 M-type stars with 26,159 flares (48\%). These spectral-type-dependent values are broadly consistent with the overall predicted CME occurrence fraction.

To assess the RF model-predicted CME association rates for flares on FGKM-type main-sequence stars, we compared our results with existing large-sample time-domain spectroscopic surveys of stellar CMEs. \citet{2024arXiv240711461V} conducted 10-day time-resolved H$\alpha$ spectroscopic monitoring of 22 stars---10 in the Praesepe cluster and 12 in the Pleiades cluster---but did not detect any H$\alpha$ wing asymmetries attributable to stellar CME eruptions. Using the flare-search procedure developed in this work, we analyzed the 2-minute cadence \textit{TESS} light curves of these 22 stars. The Pleiades targets were observed in Sectors 42, 43, 44, 70, and 71, while the Praesepe targets were covered by Sectors 44, 45, 46, and 72. Summing the observation durations for each star yields a total \textit{TESS} baseline of approximately 2083~days, during which we identified 136 flares, corresponding to an average flare occurrence rate of about $0.65$ flares per 10~days. Applying our trained model to these flares indicates that 72 events are predicted to be associated with CMEs, giving a CME association fraction of $53\%$, which is $6\%$ higher than the overall CME association probability for \textit{TESS}-detected flares on FGKM-type main-sequence stars ($47\%$). These results suggest that selecting these 22 stars as long-term spectroscopic targets for CME searches is reasonable. The nondetection of CMEs in the 10-day monitoring of \citet{2024arXiv240711461V} is likely related to the intrinsically low flare occurrence rate ($\sim 0.65$ flares per 10~days), and continued monitoring would substantially increase the probability of detecting stellar CMEs in this sample. Under the assumption that flare occurrences follow a Poisson process, and adopting the previously estimated mean rate of $\sim 0.65$ flares per 10 days, the theoretical probability of detecting no flares within a 10-day observing window is approximately $52\%$.

In addition, \citet{2019A&A...623A..49V} reported 478 spectra with Balmer-line profile asymmetries out of a total of 5500 spectra, and identified 25 stars exhibiting relatively high CME occurrence rates (1.2--19.6 events per day). These 25 stars are covered by 58 TESS sectors. The combined TESS time baseline for these stars is approximately 2898~days, within which we detected a total of 2855 flares, corresponding to an average flare occurrence rate of about $1.0$ flare per day. Applying the RF model to these flares yields 1424 events predicted to be accompanied by CMEs, corresponding to a CME association fraction of $\sim 49\%$. This value is slightly higher than the overall $47\%$ CME association probability derived for FGKM-type main-sequence stars, and is broadly consistent with the high CME occurrence rates reported by \citet{2019A&A...623A..49V} for these 25 targets.

Considering that the TESS stellar white-light flares are sampled at a cadence of 2 minutes, whereas the GOES soft X-ray flares used for model training have a cadence of 1 minute, we tested whether this difference in temporal resolution affects the prediction results. To do so, we linearly downsampled the solar soft X-ray light curves to match the 2-minute cadence of the TESS data, and then applied the same preprocessing and feature-extraction procedures. Using the best-performing random forest classifier, we retrained and re-evaluated the model on the downsampled solar dataset. The resulting accuracy, F1 score, and TSS show only negligible differences compared to the original 1-minute–cadence results. We then used this newly trained classifier to predict CME associations for the TESS stellar flare sample. The predicted CME fraction is 46\%, essentially identical to the original value of 47\%. These tests demonstrate that the difference between 1-minute and 2-minute sampling has a negligible impact on model performance or prediction outcomes. Therefore, we base the subsequent analysis on the original 1-minute–cadence GOES soft X-ray flare data.

\subsection{Distribution of Model-predicted CMEs with Flare Energy}
In this work, the bolometric energy of each detected flare is estimated following the formulation provided by \citet{2024MNRAS.527.8290P}:
\begin{equation}
E_{\text{flare}} = E_{\text{bol}} = \frac{ED \times L_{\text{TESS}}}{c}
\end{equation}
where $ED$ is the equivalent duration of the flare, $L_{\text{TESS}}$ is the quiescent stellar luminosity in the TESS bandpass, and $c$ is a photometric calibration factor that converts the observed TESS-band flare enhancement into bolometric flare energy. The equivalent duration is defined as
\begin{equation}
ED = \int \frac{F_{\rm flare}(t) - F_{\rm quiescent}}{F_{\rm quiescent}} \, dt,
\end{equation}
which represents the amount of time the quiescent star would need to radiate the same TESS-band energy as released during the flare. To derive $L_{\text{TESS}}$, we use the TESS zero-point fluxes and magnitude relations provided by \citet{2015ApJ...809...77S}, together with standard luminosity--flux--distance conversions. We adopt the calibration factor $c = 0.19$, as recommended by \citep{2022ApJ...926..204H} and \citep{2024MNRAS.527.8290P}, corresponding to an assumed flare blackbody temperature of approximately 9000~K. This temperature is widely used in studies of solar and stellar white-light flares: solar white-light flare continua are broadly consistent with a $\sim$9000~K blackbody \citep{2011A&A...530A..84K}, and M-dwarf flare analyses show typical effective blackbody temperatures of 9000--10000~K \citep{2022A&A...668A.111M}. Although exceptionally energetic superflares may exhibit transient peak temperatures up to $\sim$30000~K \citep{2020ApJ...902..115H}, using 9000~K as an average flare temperature provides a reasonable estimate of the total flare energy over the event duration. Based on this, we adopt 9000~K as the representative flare blackbody temperature in this study.

Following the above calculations, flare energies were derived for all samples, ranging from $4.45 \times 10^{30}$ to $1.28 \times 10^{38}$ erg. Notably, only two flares exhibit energies near $10^{30}$ erg, making this range statistically negligible. Consequently, nearly all flare energies are distributed between $10^{31}$ and $10^{38}$ erg, consistent with the energy distribution reported by previous stellar flare studies \citep[e.g.,][]{2012Natur.485..478M, 2019ApJS..243...28L, 2024Sci...386.1301V, 2025ApJS..276...44S}.

Using the calculated flare energies and stellar classifications, We examined the distribution of eruptive flares predicted by the RF model across stellar types, as well as their relative fractions in different energy bins. Observational statistics of solar flares demonstrate that the likelihood of CME association increases with flare energy; however, our data do not exhibit such a trend. Previous solar studies have shown that CME association increases with flare intensity. For example, \citet{2009IAUS..257..233Y} reported CME association rates of approximately 20\% for C-class, 49\% for M-class, and 91\% for X-class flares. \citet{2011SoPh..268..195A} further identified a roughly logarithmic relationship between CME mass and flare peak radiation, implying that more energetic flares tend to produce more massive CMEs.

\begin{figure*}[ht!]
\plotone{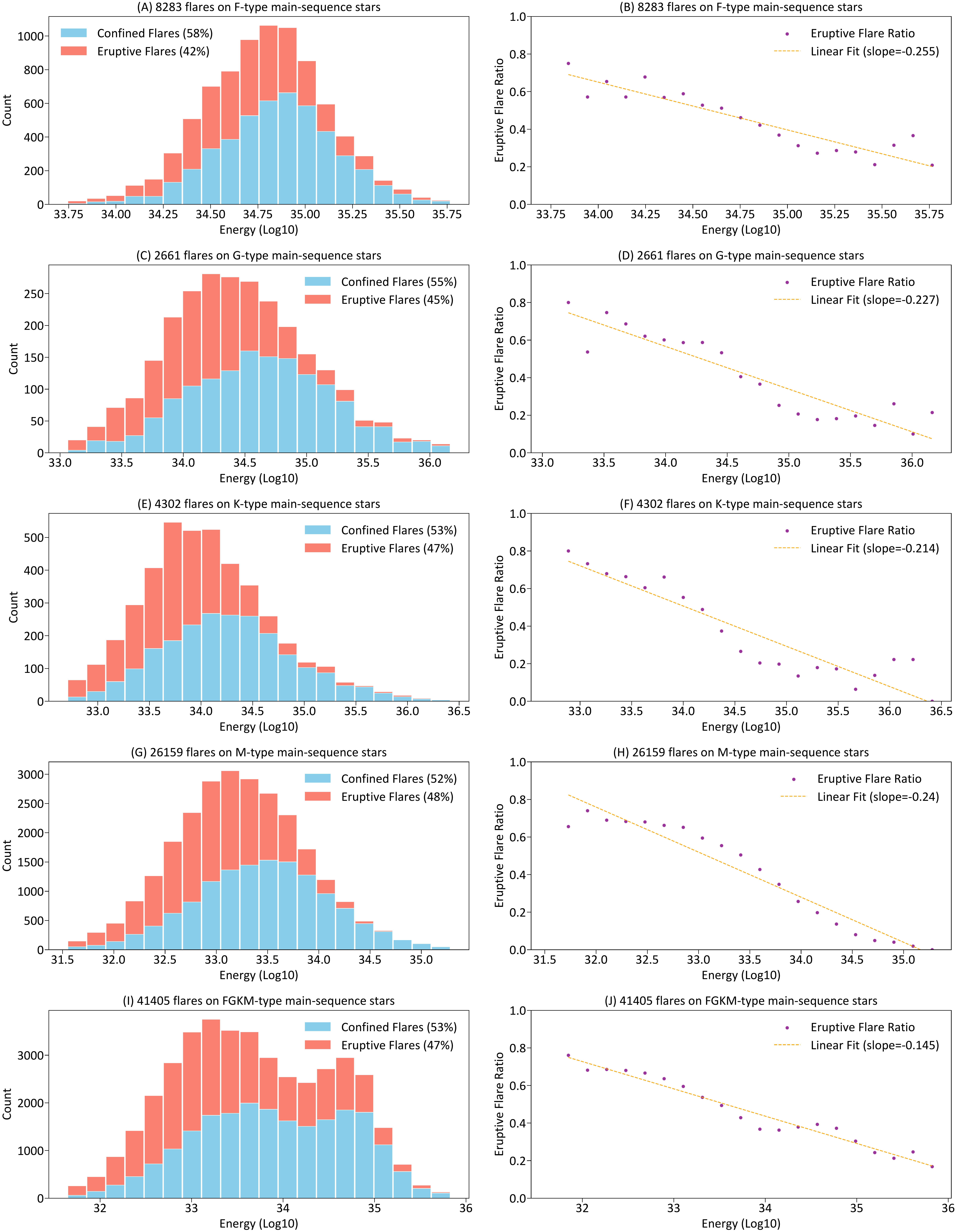}
\caption{Distribution of the model-predicted eruptive-flare fraction as a function of flare energy for FGKM-type main-sequence stars at a flare blackbody temperature of 9000~K. Panels (A)–(H) correspond to F-, G-, K-, and M-type stars, with two panels for each spectral type. The left panels show flare-count histograms, separating confined flares (blue) and eruptive flares (orange), with the corresponding eruptive fractions indicated in the legends. The right panels display the eruptive-flare fraction as a function of flare energy, together with a linear-fit trend (orange dashed line). Panels (I) and (J) show the combined results for all FGKM-type stars. Overall, the eruptive-flare fraction decreases with increasing flare energy.}
\label{fig:fig3}
\end{figure*}

\begin{figure*}[ht!]
\plotone{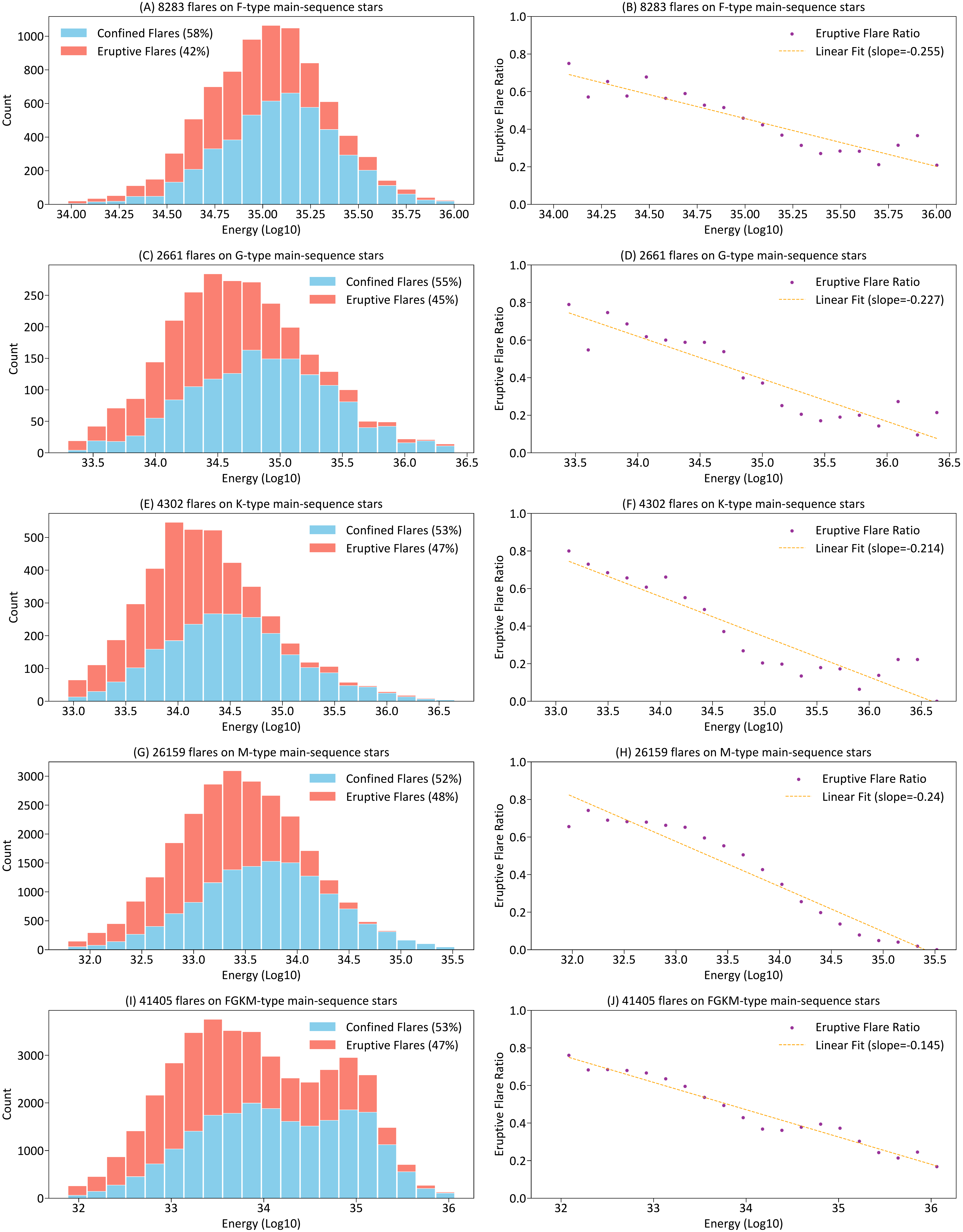}
\caption{Distribution of the model-predicted eruptive-flare fraction as a function of flare energy for FGKM-type main-sequence stars at a flare blackbody temperature of 12000~K. Panels (A)–(H) correspond to F-, G-, K-, and M-type stars, with two panels for each spectral type. The left panels show flare-count histograms, separating confined flares (blue) and eruptive flares (orange), with the corresponding eruptive fractions indicated in the legends. The right panels display the eruptive-flare fraction as a function of flare energy, together with a linear-fit trend (orange dashed line). Panels (I) and (J) show the combined results for all FGKM-type stars. Overall, the eruptive-flare fraction decreases with increasing flare energy.}
\label{fig:fig4}
\end{figure*}

\begin{figure*}[ht!]
\plotone{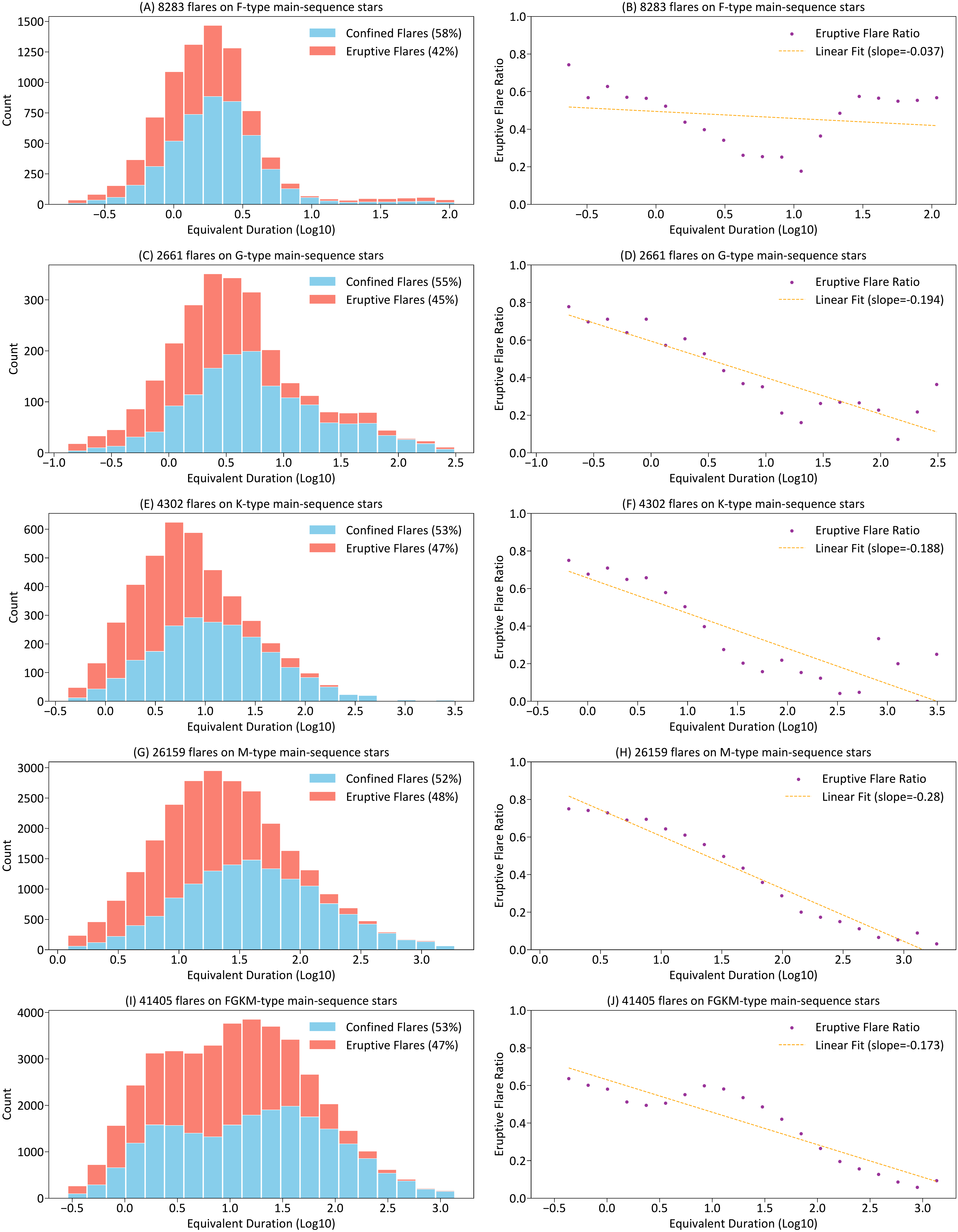}
\caption{Distribution of the model-predicted eruptive-flare fraction as a function of flare equivalent duration (ED) for FGKM-type main-sequence stars. Panels (A)–(H) correspond to F-, G-, K-, and M-type stars, with two panels for each spectral type. The left panels show flare-count histograms, separating confined flares (blue) and eruptive flares (orange), with the corresponding eruptive fractions indicated in the legends. The right panels display the eruptive-flare fraction as a function of flare ED, together with a linear-fit trend (orange dashed line). Panels (I) and (J) show the combined results for all FGKM-type stars. The eruptive-flare fraction decreases with increasing flare ED.}
\label{fig:fig5}
\end{figure*}

\begin{figure*}[ht!]
\plotone{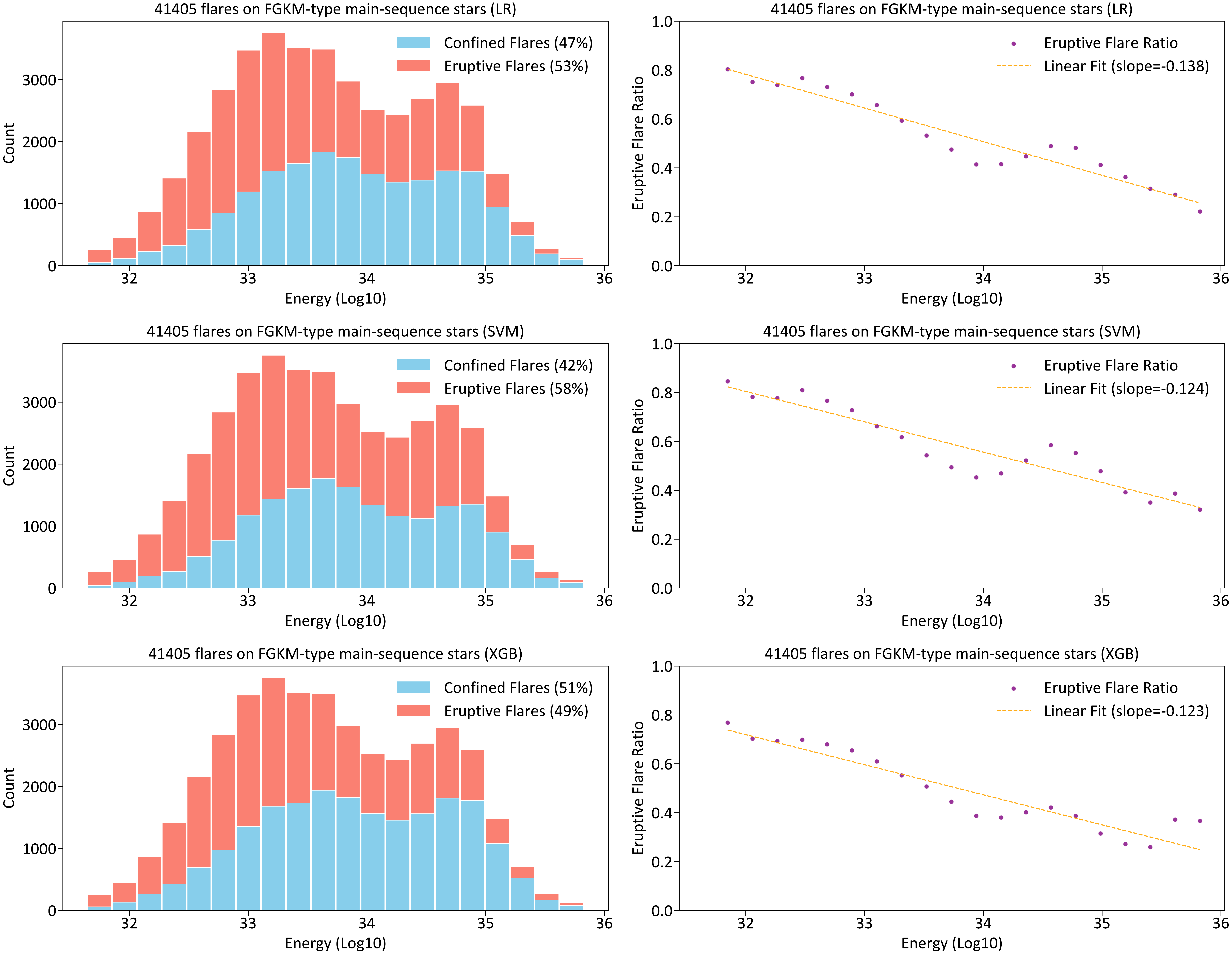}
\caption{Distribution of the model-predicted eruptive-flare fraction as a function of flare energy for FGKM-type main-sequence stars at a flare blackbody temperature of 9000~K, based on the LR, SVM, and XGB models. The left panel in each row shows the flare-count histogram, separating confined flares (blue) and eruptive flares (orange), with the corresponding eruptive fraction indicated in the legend. The right panel displays the overall eruptive-flare fraction as a function of flare energy, along with a linear-fit trend (orange dashed line). Overall, the eruptive-flare fraction decreases with increasing flare energy.}
\label{fig:fig6}
\end{figure*}

\begin{figure*}[ht!]
\plotone{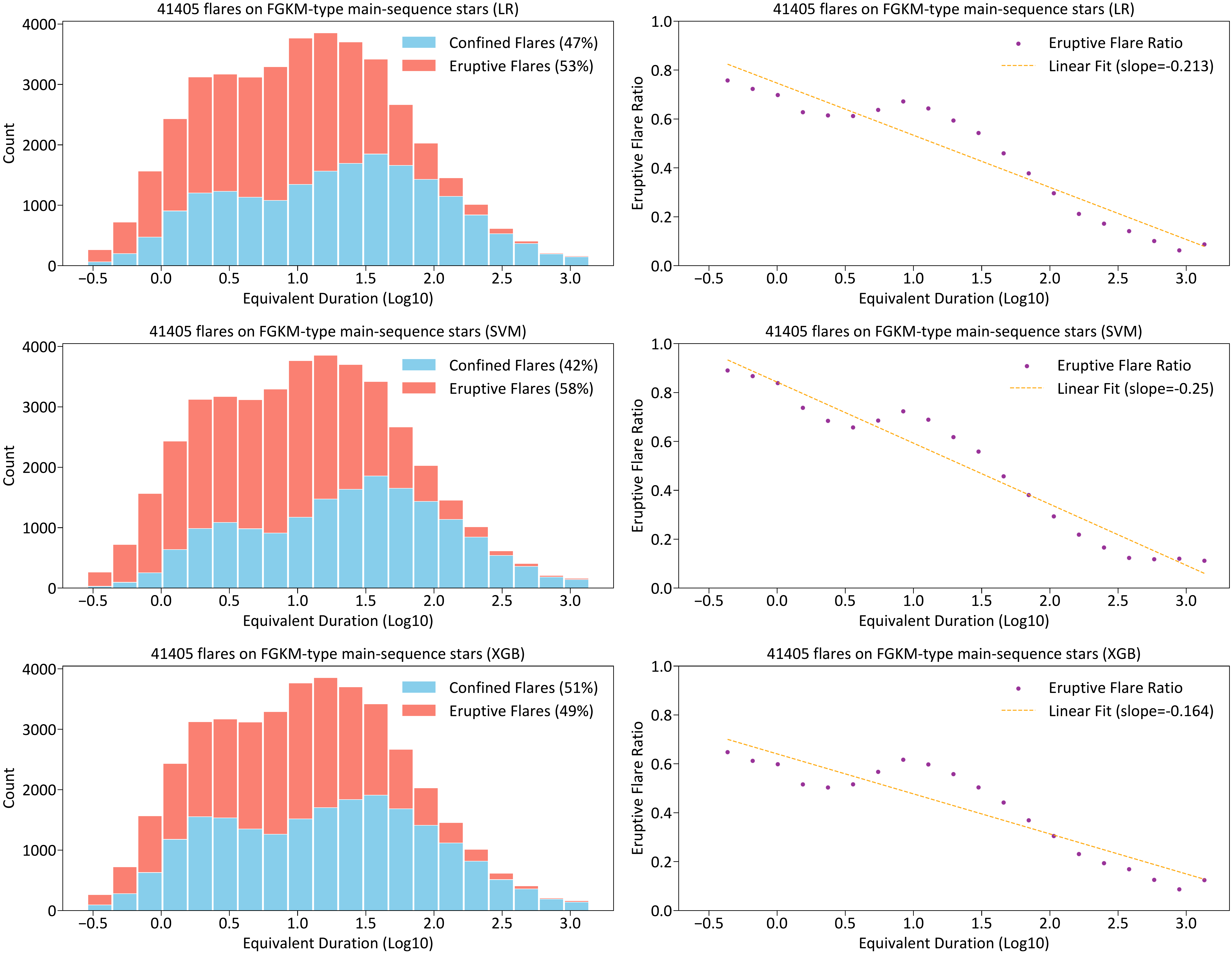}
\caption{Distribution of the model-predicted eruptive-flare fraction as a function of flare equivalent duration (ED) for FGKM-type main-sequence stars, based on the LR, SVM, and XGB models. The left panel in each row shows the flare-count histogram, separating confined flares (blue) and eruptive flares (orange), with the corresponding eruptive fraction indicated in the legend. The right panel displays the overall eruptive-flare fraction as a function of flare ED, along with a linear-fit trend (orange dashed line). Overall, the eruptive-flare fraction decreases with increasing flare ED.}
\label{fig:fig7}
\end{figure*}

Figure~\ref{fig:fig3} shows the fraction of eruptive flares predicted by the RF model as a function of flare energy for FGKM-type main-sequence stars. It can be seen that the fraction of eruptive flares systematically decreases with increasing flare energy, indicating that statistically, more energetic flares are less likely to be associated with CMEs. This trend is consistent across all spectral types, suggesting that it is not driven by statistical uncertainties or sample bias. To examine whether this behavior depends on uncertainties in the flare energy estimates, Figure~\ref{fig:fig4} presents the RF model-predicted eruptive flare fractions assuming a flare blackbody temperature of 12,000\,K, while Figure~\ref{fig:fig5} shows the predicted fractions as a function of equivalent duration. Figures~\ref{fig:fig6} and \ref{fig:fig7} further display the overall flare--CME association rates predicted by the LR, SVM, and XGB models as functions of flare energy and equivalent duration. All models show a similar decreasing trend in the fraction of eruptive flares. These results indicate that the declining trend is robust under different single-temperature assumptions. Overall, the predicted CME association rate decreases with increasing flare energy and equivalent duration, implying that flares with stronger white-light enhancements tend to have lower CME association probabilities.

These results rely on extrapolating the solar flare–CME relationship to the more energetic flares observed on active FGKM-type main-sequence stars. The validity of such an extrapolation is supported by several observational and theoretical studies. Space-based photometry from Kepler and TESS has detected superflares on Sun-like stars with energies far exceeding those of modern solar events, yet their flare frequency–energy distributions exhibit power-law indices similar to the solar case \citep[e.g.,][]{2015EP&S...67...59M}. This suggests that solar and stellar superflares are governed by a common magnetic-reconnection mechanism \citep[e.g.,][]{2012Natur.485..478M, 2024Sci...386.1301V}, with the primary difference arising from the larger magnetic fluxes or magnetic-field scales involved \citep[e.g.,][]{2013ApJ...771..127N, 2020Sci...369..694Y, 2024Sci...386...76Y}. This provides provide a physical basis for extending solar flare–CME relations to more energetic stellar regimes.

On the Sun, stronger flares tend to exhibit higher CME association rates \citep[e.g.,][]{2009IAUS..257..233Y}. However, recent statistical studies show that this relationship depends sensitively on the evolutionary stage of the active region. During the developing phase—when background fields are stronger and magnetic complexity increases—flare occurrence is enhanced, but the fraction of eruptive events remains low. In contrast, during the decaying phase—when overlying fields weaken—the probability of producing an eruptive flare increases \citep{2024ApJ...976L...2L}. These findings highlight the crucial role of the background magnetic field in determining whether a flare becomes eruptive or confined.

This physical picture is consistent with numerical MHD simulations showing that large-scale overlying magnetic fields on active stars can strongly suppress CMEs, with dipolar field strengths of several tens to over one hundred gauss sufficient to trap otherwise solar-like eruptions unless the release energy is extremely high \citep[e.g.,][]{2018ApJ...862...93A}. From the perspective of instability criteria, strong overlying fields above large starspot regions produce broader torus-stable zones, thereby raising the threshold for flux-rope destabilization and lowering the likelihood of successful eruptions \citep[e.g.,][]{2022MNRAS.509.5075S}. Since the majority of TESS-detected flares occur on rapidly rotating, magnetically active stars with large starspots, these considerations naturally imply that not all stellar flares are accompanied by CMEs. More energetic flares are typically associated with stronger background magnetic fields, which may further enhance confinement and reduce the eruptive fraction. Taken together, these studies provide physical support for the reliability of our model predictions.

Moreover, a potential explanation for the decreasing CME fraction with increasing flare energy in the TESS sample is observational selection effects arising from wavelength‐dependent detectability and instrumental sensitivity. Our model is trained on GOES soft X‐ray flares, whereas the stellar predictions rely on TESS white‐light flares, and these two bands do not trace flare energetics in the same way: on the Sun, many strong soft X‐ray flares show little or no white‐light enhancement, and white-light events represent only a small, energetically extreme subset of all flares \citep{1989SoPh..121..261N, 2014ApJ...783...98K}. Moreover, TESS’s red-optical band and limited photometric sensitivity bias detections toward large-amplitude, long-duration flares, while numerous smaller or more impulsive events—some of which might be CME-producing—fall below the detection threshold \citep{2020AJ....159...60G}. These combined effects imply that the apparent decrease in CME fraction with flare energy may partly reflect band-pass and sensitivity‐driven selection biases rather than intrinsic differences in flare–CME coupling. Simultaneous X‐ray and optical monitoring will be essential for establishing the true CME–flare energy relationship on FGKM-type main-sequence stars.

\section{Summary} \label{sec:summy}
We constructed a training dataset from the GOES XRS 1--8~\AA\ solar flare light curves, extracting and normalizing features using a standardized feature extraction pipeline. Five widely used machine learning classifiers were trained and evaluated, among which the Random Forest algorithm was selected as the optimal model for stellar flare prediction. Subsequently, we searched for flares in TESS light curves and manually vetted the candidates to compile a high-quality sample. The same processing and normalization procedures as used for the solar flares were applied to each stellar flare.

We subsequently trained classification models using the solar flare dataset. To evaluate model performance under class imbalance, we employed the TSS, which provides a balanced assessment regardless of sample distribution. Among the tested models, the Random Forest classifier consistently outperformed others across evaluation metrics and exhibited the most stable performance. Based on these results, it was selected as the final model for predicting CME occurrence in stellar flares.

To assess the generalizability of the solar-flare-trained model to stellar samples, we first compared the light curve morphologies of solar flares observed in the GOES XRS 1--8~\AA\ band with white-light flares. After uniform data preprocessing, the two flare types exhibit highly similar temporal evolution, providing empirical support for applying the solar-trained model to stellar white-light flares. Based on this, we classified the stellar spectral types following the method of \citet{2018A&A...616A..10G} and applied the RF model, which shows the best overall performance, to all 41,405 TESS-detected main-sequence white-light flares to predict their CME association. The results indicate that approximately 47\% of the flare events exhibit CME-like morphological features, with the model-inferred intrinsic CME association fraction ranging from 35\% to 60\%. When further subdivided by spectral type, the predicted CME association fraction for FGKM main-sequence stars remains roughly consistent at $\sim$47\%. 

Based on Figure~\ref{fig:fig3}, we analyzed the distribution of eruptive flares across different energy ranges. The results reveal a decreasing proportion of eruptive flares with increasing flare energy. This trend may be explained by differences in stellar magnetic field strength. Stars with stronger magnetic fields tend to produce more energetic superflares; however, their intense magnetic confinement can inhibit efficient plasma ejection, thereby reducing the likelihood of CME occurrence. Thus, higher flare energy does not necessarily correspond to a higher CME rate. These findings are consistent with the results of \citet{2024ApJ...976L...2L} and \citet{2022MNRAS.509.5075S}, and are supported by our analysis of both the overall sample and the main-sequence subsamples across different spectral types.

The proposed RF model demonstrates preliminary efficacy in predicting stellar CMEs but is limited by several factors. The training data consist primarily of solar M- and X-class flares observed in the soft X-ray band rather than the scarce white-light flares, which could introduce potential physical inconsistencies. The solar training dataset is limited in sample size, which constrains the diversity of flare morphologies available to the model; future expansions of Sun-as-a-star flare observations will provide larger and more representative datasets for further improving model training. This model achieved a TSS of 0.31, demonstrating the feasibility of transfer learning from solar to stellar flares using morphological features to predict CME occurrence. Feature importance analysis highlights the dominant contribution of image features extracted by ResNet50; however, the inherent “black-box” nature of the network constrains physical interpretability. Incorporation of multi-band, multi-timescale observations in future work is expected to enhance model robustness and interpretability. This framework lays a scalable foundation for statistical modeling of stellar CMEs and exoplanet space weather assessment.

\begin{acknowledgments}
We gratefully acknowledge Prof. Hui Tian, Prof. Ying Li, Dr. Yijun Hou, Dr. Hechao Chen, and Dr. Lei Zhang for their insightful comments and constructive suggestions that greatly improved this study. This work is supported by the National Natural Science Foundation of China (NSFC) under grant 12473055, 12250006, and by the Guizhou University Natural Science Special Research Fund under project number 202358. It is further supported by the Guizhou Provincial Science and Technology Projects (grant No. QianKeHe-JiChu-MS[2025]695). We acknowledge the use of soft X-ray flare observations from GOES, with data publicly available from the National Centers for Environmental Information (NCEI) at https://www.ngdc.noaa.gov/stp/satellite/goes/. We also acknowledge the use of data from TESS, obtained from the MAST archive at the Space Telescope Science Institute (STScI). Funding for the TESS mission has been provided by the NASA Explorer Program. STScI is operated by the Association of Universities for Research in Astronomy, Inc., under NASA contract NAS 5–26555. This research made use of the Python programming language and the scientific packages NumPy, SciPy, scikit-learn, and Matplotlib. This research made use of the Python programming language and the scientific packages NumPy \citep{2020Natur.585..357H}, SciPy \citep{2020NatMe..17..261V}, scikit-learn \citep{2011JMLR...12.2825P}, and Matplotlib \citep{2007CSE.....9...90H}.
\end{acknowledgments}

\appendix

\counterwithin{equation}{section}
\setcounter{figure}{0}
\renewcommand{\thefigure}{A\arabic{figure}}
\renewcommand{\theHfigure}{A\arabic{figure}}
\renewcommand{\theequation}{\thesection\arabic{equation}}
\renewcommand{\thetable}{A\arabic{table}}
\renewcommand{\theHtable}{A\arabic{table}}
\setcounter{table}{0}

\section{Machine Learning Models} \label{sec:A}
\subsection{Linear Models}
Support vector machines (SVMs) are supervised classifiers that construct optimal hyperplanes by maximizing class margins \citep{1995MachL..20..273C}. For linearly or nearly separable data, SVMs improve generalization by maximizing the margin between classes in the feature space. In this study, we use a linear SVM due to its robustness and computational efficiency in high-dimensional feature spaces.

Logistic regression (LR) is a generalized linear model commonly used for binary classification. It estimates class probabilities by applying a sigmoid function to a linear combination of features, with parameters fitted via maximum likelihood \citep{1958JRSB...20..215C}. LR is favored for its simplicity, interpretability, and suitability for problems with moderate dimensionality and approximate linearity.

\subsection{Discriminant Model}
Linear discriminant analysis (LDA) is a classical supervised technique for dimensionality reduction and classification. It finds a linear projection that maximizes between-class variance and minimizes within-class variance. This approach assumes classes follow multivariate normal distributions with a common covariance matrix. Proposed by \citet{1936AnEug..7..179F}, LDA is widely used in pattern recognition and classification.

\subsection{Ensemble Models}
Random Forest (RF) is an ensemble method that uses multiple decision trees to improve model stability and generalization. It generates bootstrap samples of the data to train individual trees independently. Final predictions are obtained via majority voting \citep{2001MachL..45....5B}. At each split, RF randomly selects a subset of features to reduce overfitting.

Extreme Gradient Boosting (XGBoost) is an efficient ensemble algorithm introduced by \citet{2016KDD...785-794C}. It extends traditional gradient boosting by incorporating second-order derivatives, regularization, and reduced learning rates, thereby enhancing model expressiveness and mitigating overfitting. XGBoost also supports parallel computing and built-in cross-validation, improving training efficiency and stability.

\subsection{Baseline Model}
A dummy classifier serves as a baseline that does not use input features; its predictions rely solely on the prior distribution of the target variable in the training set. For binary classification, it typically adopts a majority-class prediction strategy. It requires no parameter training and is commonly used as a benchmark to assess the added value of models utilizing input features.

\section{Evaluation Metrics} \label{sec:B}
\setcounter{equation}{0}
The confusion matrix is a fundamental tool for evaluating classification performance and comprises true positives (TP), true negatives (TN), false positives (FP), and false negatives (FN). $TP$ and $TN$ denote correctly classified outburst and non-outburst events, respectively; $FP$ and $FN$ represent misclassified cases. These components form the basis of key evaluation metrics introduced in the following section.

$Accuracy$ indicates the proportion of the model’s total predictions that are correct. It is defined as follows:
\begin{equation}
Accuracy = \frac{TP + TN}{TP + TN + FP + FN}
\label{eq:accuracy}
\end{equation}

$Precision$ measures the proportion of all events predicted as outbreaks that are truly outbreaks. It is defined as:
\begin{equation}
Precision = \frac{TP}{TP + FP}
\label{eq:precision}
\end{equation}

The $F1$ index is the harmonic mean of precision and recall, providing a balanced measure of model performance without favoring either metric.
\begin{equation}
F_1 = \frac{2TP}{2TP + FP + FN}
\label{eq:f1_score}
\end{equation}

The false alarm rate (FAR) represents the proportion of events incorrectly classified as outbreaks despite being non-outbreak events. In this paper, FAR is used interchangeably with the conventional false positive rate (FPR), defined as follows.
\begin{equation}
FAR = \frac{FP}{FP + TN}
\label{eq:far}
\end{equation}

The false discovery rate (FDR) quantifies the proportion of predicted outbreaks that are actually false positives (FPs) and is defined as follows.
\begin{equation}
FDR = \frac{FP}{FP + TP}
\label{eq:fdr}
\end{equation}

True Skill Statistic (TSS) measures a model’s ability to accurately identify both outbreak and non-outbreak events, independent of class imbalance. It is a robust metric commonly used for predicting extreme events and is defined as:
\begin{equation}
TSS = \frac{TP}{TP + FN} - \frac{FP}{FP + TN}
\label{eq:tss}
\end{equation}

A robust classification model should minimize both the false alarm rate (FAR) and false discovery rate (FDR); however, improving one often compromises the other. In this study, the true skill statistic (TSS), which is insensitive to class imbalance, is adopted as the primary evaluation metric. Additionally, conventional metrics such as $Accuracy$, $Precision$, and $F1$ score are employed for a comprehensive assessment.

\section{Stellar Flare Search} \label{sec:C}
This study identifies flare events using stellar light curves with a 2-minute cadence from TESS sectors 1 to 81. However, the raw light curves are contaminated by various noise sources and are therefore unsuitable for direct flare detection, necessitating uniform preprocessing before formal analysis.

\begin{figure*}[ht!]
\plotone{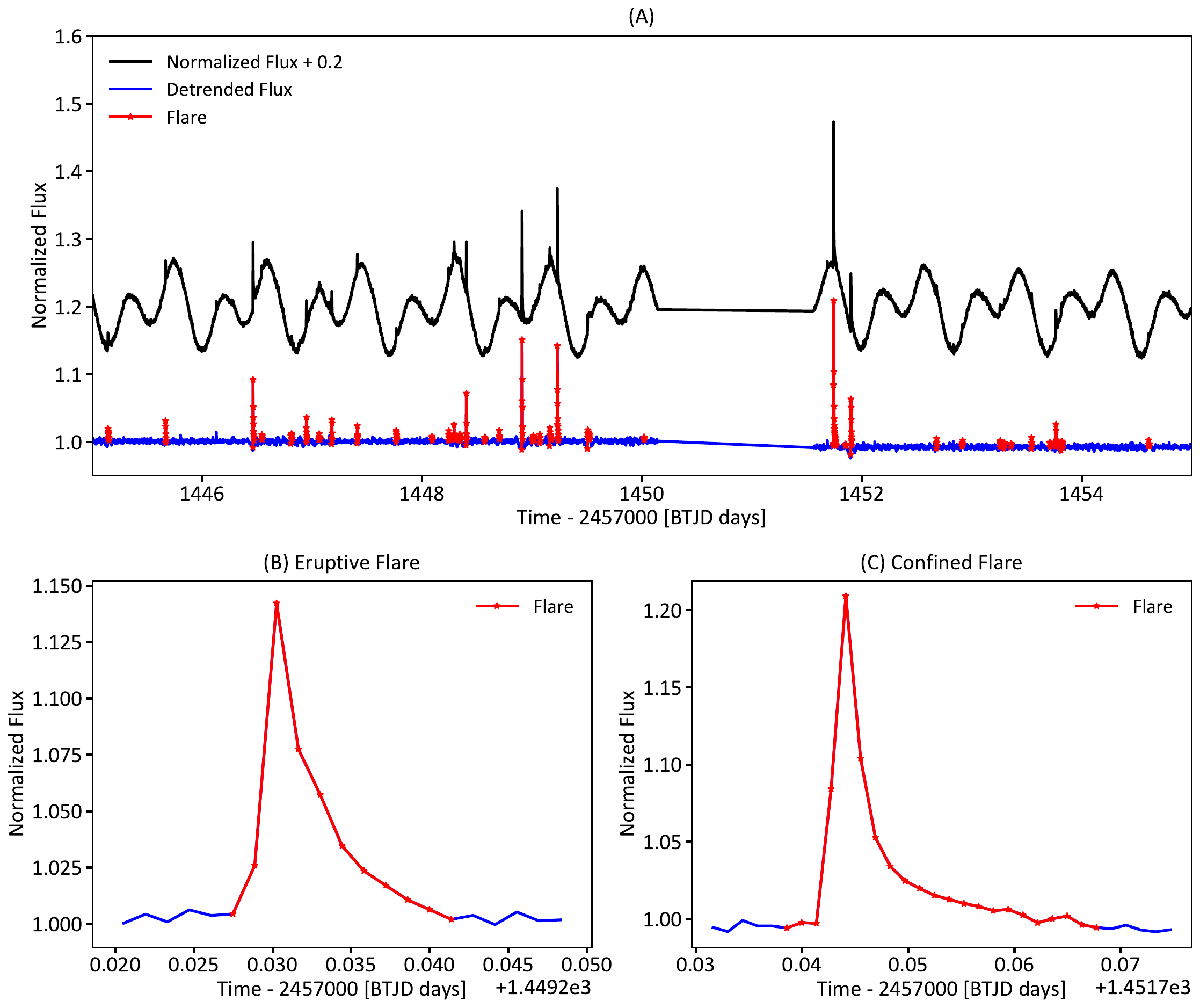}
\caption{Light curve and flare features of TIC 220433364 in Sector 5. Panel (A) shows the full light curve, including PDCSAP flux (black; offset by +0.2), detrended flux (blue), and identified flares (red markers). The x-axis represents time in Barycentric TESS Julian Date (BTJD). Panels (B) and (C) provide zoomed-in views of representative eruptive and confined flares, respectively, illustrating the normalized flux over time and clearly revealing the rise, peak, and decay phases.}
\label{fig:A1}
\end{figure*}

First, each light curve is normalized using the following equation:
\begin{equation}
F_{\text{norm}} = \frac{F_{\text{PDC}} - F_{\text{med}}}{F_{\text{med}}}
\label{eq:f_norm}
\end{equation} To remove long-term background variations and normalize for differing stellar luminosities, we applied the detrending method of \citet{2022MNRAS.513.4579I} to enhance flare detectability. Following this preprocessing, flare candidates were identified and extracted from the light curves. For example, Figure~\ref{fig:A1} shows the light curve of TIC 220433364 (Sector 5). Each candidate segment includes the flare core plus five data points before and after, with corresponding images saved to assist background subtraction.

Based on the preliminary search results, we developed an automated screening procedure to enhance flare identification accuracy and eliminate false signals. This procedure primarily consists of the following criteria:
\begin{enumerate}[label=(\arabic*)]

    \item Peak Structure Constraint: To identify flare events in stellar light curves, we used the flare detection method implemented in the altaipony package. Specifically, a flare is defined as at least three consecutive data points with residual flux significantly exceeding the background (default threshold: three times the local standard deviation), remaining statistically significant after accounting for measurement errors. This algorithm improves detection robustness by preventing noise or isolated spikes from being misclassified as genuine flares.

    \item Temporal Structure Constraint: The number of data points in the flare’s rise phase must be fewer than that in the decay phase, and the entire flare region must contain at least four valid data points.

    \item Linear Fit Constraint: Linear fits were performed separately on the rise and decay phases, requiring a Pearson correlation coefficient of at least 0.65 to ensure well-defined flare profiles with minimal fluctuations. Additionally, the absolute slope of the rise phase must exceed that of the decay phase, consistent with the typical flare characteristic of a rapid rise followed by a gradual decay.
    
\end{enumerate}

Applying the above selection criteria, we initially identified approximately 110,000 flare events. Although the automated detection process removed many false signals, a substantial number of misidentified events remained, primarily due to stellar rotation modulation, binary activity, or random noise. Such false positives are difficult to eliminate entirely through automated algorithms. To further refine the sample, we conducted manual visual inspection of each candidate against the full light curve background, removing clearly spurious or background-contaminated events. After vetting, 60,873 reliable flares were retained. However, a significant fraction originated from non-FGKM stars, which fall outside the scope of this study. Based on stellar classification (see Appendix~\ref{sec:D}), these were excluded. The final high-quality sample, listed in Table~\ref{tab:stella}, comprises 41,405 flare events and serves as the basis for subsequent statistical analysis and physical modeling.

\begin{deluxetable*}{ccccccccc}
\tabletypesize{\small}
\tablecaption{Stellar Flare Event Catalog
\label{tab:stella}}
\tablehead{
\colhead{No.} & \colhead{Star Name} & \colhead{Start Time} & \colhead{End Time} & \colhead{ED} & \colhead{Amplitude} & \colhead{Duration} & \colhead{Energy} & \colhead{CME Association}
}
\startdata
1     & TIC 25117273   & 1341.278743 & 1341.29541  & 4.8547    & 0.0057    & 0.0167    & $1.91\times10^{35}$  & 0 \\
2     & TIC 25132999   & 1340.956554 & 1341.023221 & 120.0938  & 0.0973    & 0.0667    & $4.83\times10^{34}$  & 0 \\
3     & TIC 25134376   & 1342.163376 & 1342.168931 & 9.6998    & 0.0378    & 0.0056    & $4.83\times10^{34}$  & 0 \\
4     & TIC 25153167   & 1352.378721 & 1352.391221 & 86.5595   & 0.2017    & 0.0125    & $8.09\times10^{32}$  & 1 \\
5     & TIC 25200252   & 1326.566502 & 1326.570669 & 4.5440    & 0.0198    & 0.0042    & $1.16\times10^{32}$  & 1 \\
\ldots & \ldots & \ldots & \ldots & \ldots & \ldots & \ldots & \ldots & \ldots\\
41401 & TIC 230073581 & 3512.160254 & 3512.179698 & 3.5566    & 0.0071    & 0.0194    & $1.71\times10^{33}$  & 0 \\
41402 & TIC 233738219 & 3520.326562 & 3520.357118 & 67.2421   & 0.1216    & 0.0306    & $2.36\times10^{33}$  & 0 \\
41403 & TIC 294789199 & 3525.540858 & 3525.550581 & 66.7868   & 0.2030    & 0.0097    & $4.54\times10^{33}$  & 1 \\
41404 & TIC 376013792 & 3507.343955 & 3507.355066 & 49.3653   & 0.1974    & 0.0111    & $7.60\times10^{32}$  & 1 \\
41405 & TIC 377444019 & 3521.170850 & 3521.195850 & 127.2819  & 0.1604    & 0.0250    & $1.19\times10^{33}$  & 0 \\
\enddata
\tablecomments{
Summary of the detected stellar flare parameters. Columns are: (1) sequential event index; (2) TIC identifier of the host star; (3)–(4) flare start and end times in Barycentric TESS Julian Date (BTJD = $\mathrm{BJD}_{\mathrm{TDB}} - 2457000$); (5) equivalent duration (ED) in seconds; (6) flare amplitude in relative flux units; (7) flare duration in days; (8) estimated bolometric flare energy in erg; and (9) CME association flag, where 1 denotes possible CME association and 0 indicates a confined flare. Only a subset of events is shown here for illustration. The complete table is provided in machine-readable form in the online journal.}
\end{deluxetable*}

\section{Stellar Parameters} \label{sec:D}
Stellar parameters were obtained from the TESS Input Catalog (TIC v8) via MAST using TIC IDs \citep{2019AJ....158..138S}. Extracted parameters include effective temperature, surface gravity, parallax, and Gaia G, BP, and RP magnitudes. Most values are derived from Gaia cross-matching, enabling stellar classification and further analysis.

\begin{figure*}[ht!]
\plotone{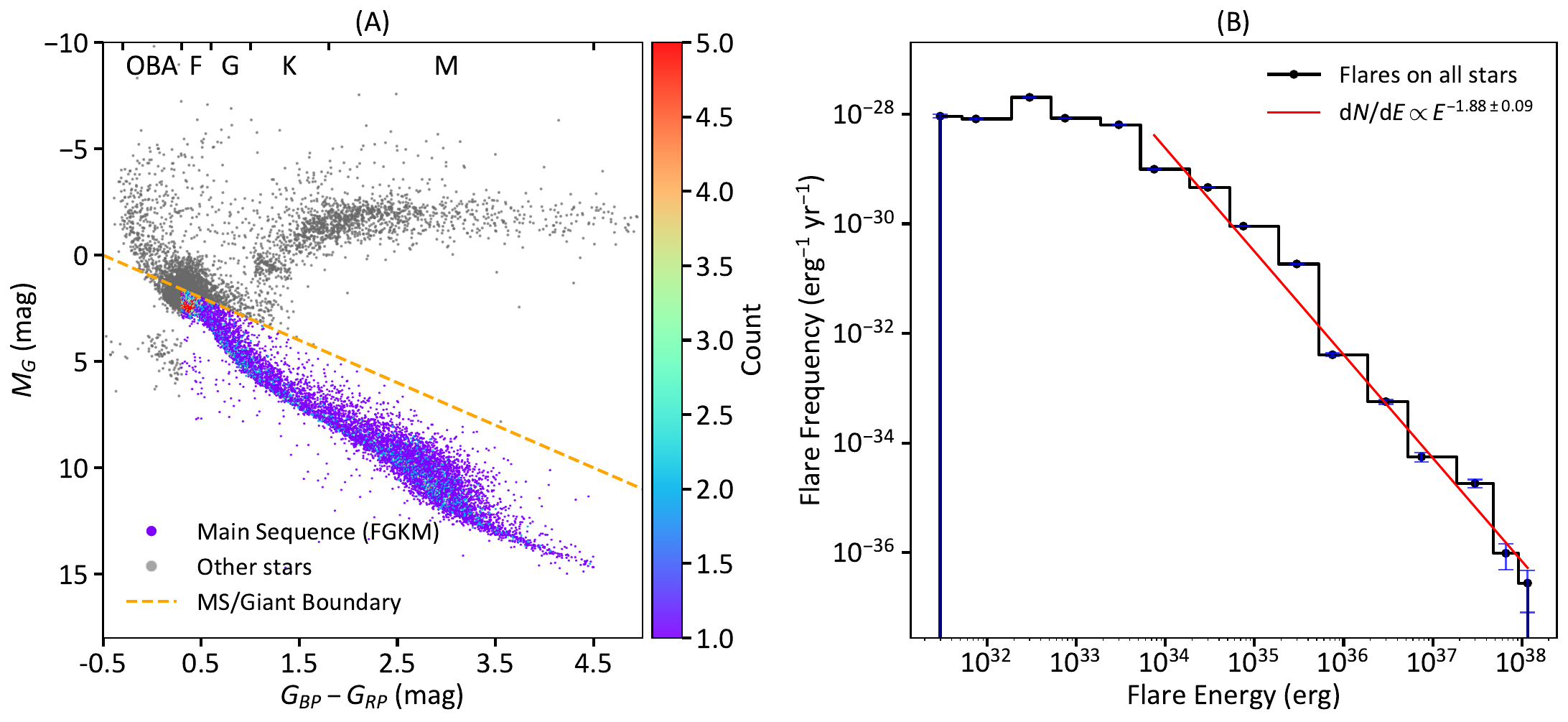}
\caption{Stellar classification and flare energy distribution. Panel (A) shows a scatter plot of absolute magnitude versus color index, where gray points above the orange dashed line denote non-FGKM main-sequence stars. The colormap shows the number density of FGKM-type main-sequence stars, outlining an HR-like structure that aids in distinguishing evolutionary stages. Panel (B) displays the flare energy–frequency distribution; observational data are shown as a black step curve with error bars, and the red solid line is a power-law fit, indicating the flare energy spectrum’s scale invariance across energy scales.}
\label{fig:A2}
\end{figure*}

To validate the incidence of stellar CMEs across different stellar types, main-sequence stars and giants are distinguished using the following empirical separation criterion:
\begin{equation}
M_G = 2 \times (G_{BP} - G_{RP}) + 1.0
\label{eq:m_g}
\end{equation} Stars below the empirical dividing line are classified as main-sequence, characterized by higher absolute magnitudes and lower luminosities, while those above are giants, exhibiting lower magnitudes and higher luminosities. Spectral types for main-sequence stars were estimated using Gaia DR2 color–magnitude distributions from \citet{2018A&A...616A..10G}. Flare and CME rates were then calculated according to stellar type. Figure~\ref{fig:A2}(A) shows that most stars are main-sequence, with only a few giants present.

Power-law distributions are characteristic of self-organized criticality (SOC) systems, where local instabilities such as magnetic reconnection can trigger large-scale energy releases \citep{1991ApJ...380L..89L}. The frequency–energy distribution of stellar flares in our sample follows a power law with an index of $1.88 \pm 0.09$ (Figure~\ref{fig:A2}(B)), which falls within the range typically reported for solar flares \citep[1.5–2.2; e.g.,][]{1993SoPh..143..275C,1995PASJ...47..251S,2002A&A...382.1070V}. Although variations in the power-law index $\alpha$ may result from differences in wavelength coverage, sample selection, or data processing methods, the broadly consistent behavior suggests a common underlying framework for magnetic activity. This supports the application of solar flare models to stellar environments and provides a basis for extending solar-based approaches to stellar systems, particularly in the context of exoplanetary space weather.

\section{Statistical Distribution and Feature Importance} \label{sec:E}
To compare manually extracted features between solar and stellar flares, we analyzed 13 parameters from both datasets. To mitigate the effect of differing sample sizes, the distributions were normalized by converting histograms into probability density functions (frequency counts divided by bin width). Extreme values were excluded by filtering data outside the 2.5th to 97.5th percentile range. The resulting density histograms reveal substantial overlaps for most features, providing a quantitative foundation to explore statistical and physical distinctions within a multidimensional parameter space.

\begin{figure*}[ht!]
\plotone{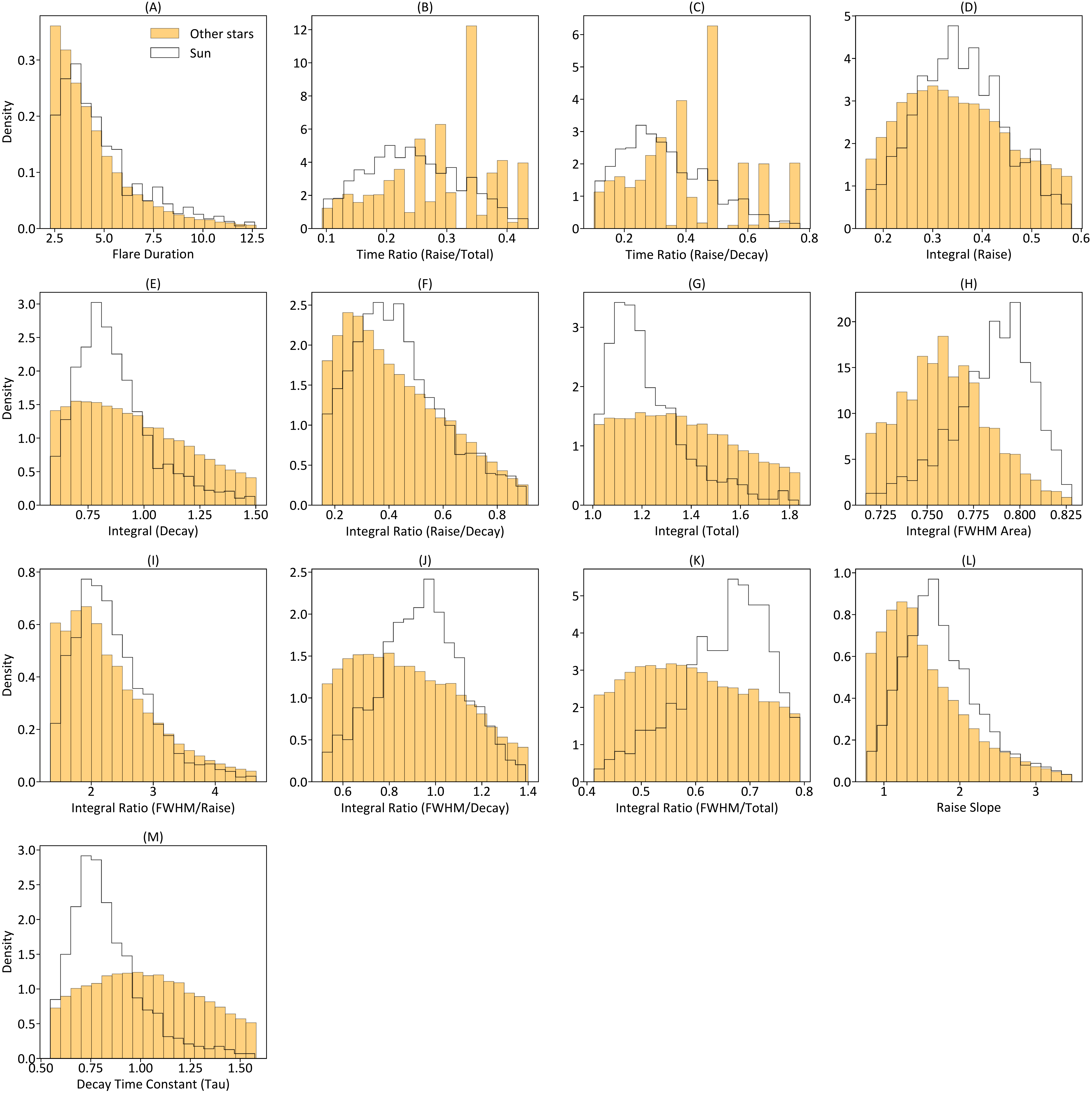}
\caption{Feature distribution maps for solar and stellar flare samples. Solar features are depicted by translucent steel-blue step histograms representing probability densities, while stellar features are illustrated by orange filled histograms representing probability densities. The distributions exhibit substantial overlap within the feature space, indicating largely shared underlying properties, though certain distinctions and partial separations are evident. This comparative analysis offers valuable insights into both the commonalities and differences of flare characteristics in solar and stellar contexts.}
\label{fig:A3}
\end{figure*}

Figure~\ref{fig:A3} shows that manually extracted features of solar and stellar flares largely overlap in the feature space, indicating similar distributions across multiple physical dimensions. To evaluate each feature’s importance in classification, we computed the Gini importance from a Random Forest (RF) model \citep{2001MachL..45....5B}, which measures the weighted decrease in Gini impurity across all tree splits. For robustness, RF models trained on the best-performing validation folds were employed. Continuous image features derived via principal component analysis (PCA) were combined into a single “image feature” to reduce dimensionality and improve interpretability.

\begin{figure*}[ht!]
\plotone{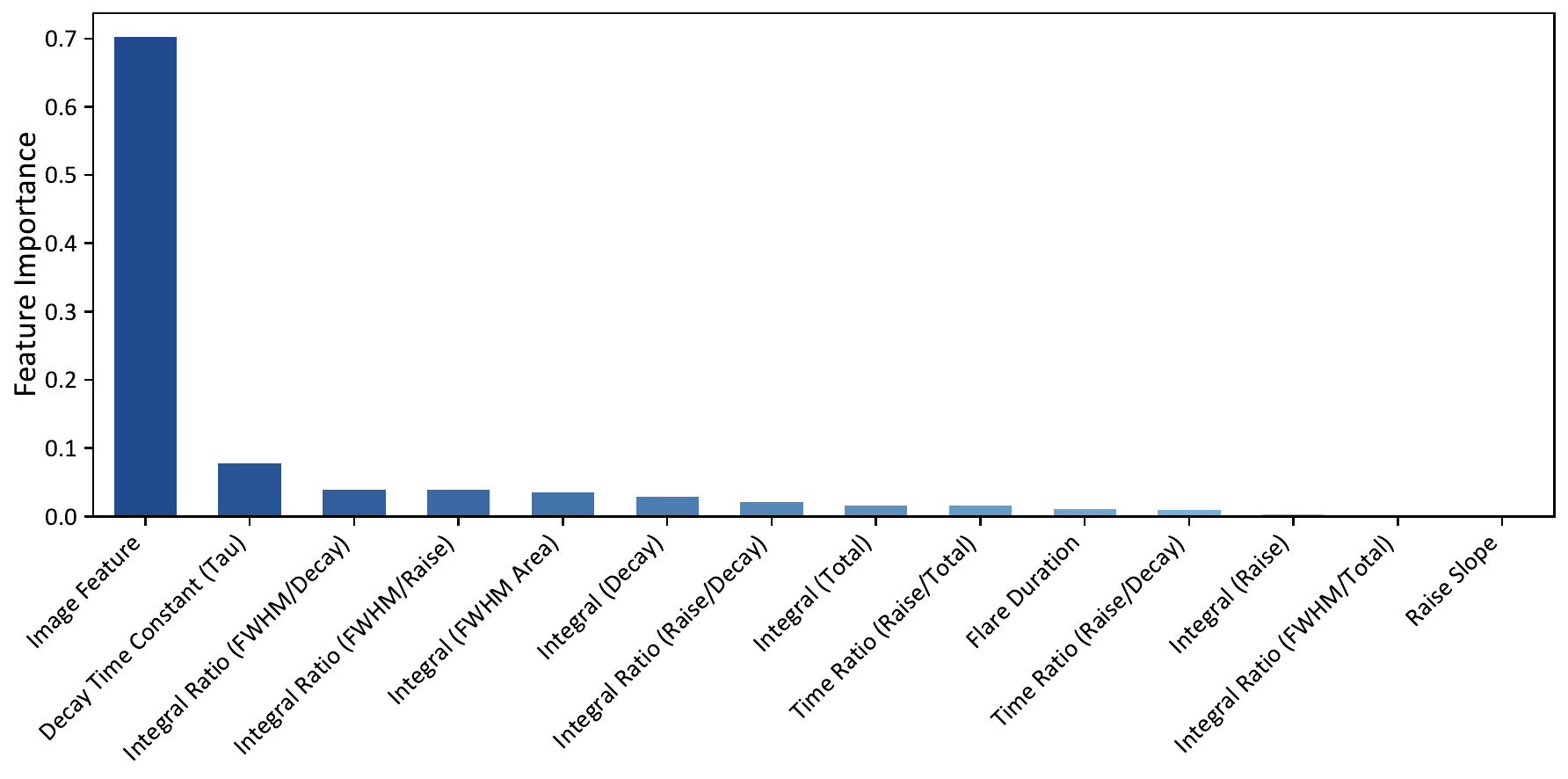}
\caption{Relative contributions of features derived from the Random Forest model. The horizontal axis lists both manually defined physical features and image-based features. The vertical axis quantifies their relative contributions to classification performance. Image features, constructed via PCA from normalized flare light curve images, form a 20-dimensional representation and dominate the contribution rankings. Their substantially higher contributions compared to individual physical parameters emphasize the strong discriminative power of deep image-derived features in flare classification.}
\label{fig:A4}
\end{figure*}

Figure~\ref{fig:A4} presents the normalized Gini importance of feature groups derived from the RF model. Image features automatically extracted by deep convolutional neural networks contribute nearly 70\% to flare classification, far surpassing manually derived physical features. This underscores the strength of deep learning in capturing complex time-domain variability. Among manual features, the most influential are the decay phase time constant, the ratios of integrals over the FWHM relative to the decay and rise phases, and the integrals of the FWHM and decay phases. These results highlight the effectiveness of a multimodal approach integrating both image-based and manually extracted features, enhancing model performance and robustness, while also guiding future feature selection and physical interpretation.

Despite their dominant contribution, image features present interpretability challenges due to the nonlinear and composite nature of deep neural networks. Nonetheless, they likely encode subtle aspects of light curve morphology, contextual structure, and noise characteristics, making them especially valuable for large-scale time-domain studies.

\section{Data Availability}
\label{sec:F}
The data and models that support the findings of this study are publicly available at \url{https://doi.org/10.5281/zenodo.15894595}. They include solar flare training and testing datasets generated from five independent random splits, the stellar flare prediction datasets along with associated host star parameter tables, sample catalog files for both solar and stellar flares, five trained Random Forest sub-models corresponding to each split, and a prediction script that performs ensemble classification and CME association on the stellar flare dataset using these sub-models.

\newpage
 
\bibliography{SY}
\bibliographystyle{aasjournalv7}



\end{document}